\documentclass[useAMS,usenatbib]{mn2e}
\usepackage{hyperref}
\usepackage{epsfig}
\usepackage{graphicx}
\usepackage{psfrag}
\usepackage{url}
\usepackage{mathrsfs}
\usepackage{subcaption}
\usepackage{multirow}

\usepackage{amsmath,amssymb}

\def\eprinttmp@#1arXiv:#2 [#3]#4@{
\ifthenelse{\equal{#3}{x}}{\href{http://arxiv.org/abs/#1}{#1}
}{\href{http://arxiv.org/abs/#2}{arXiv:#2} [#3]}}

\providecommand{\eprint}[1]{\eprinttmp@#1arXiv: [x]@}
\newcommand{\adsurl}[1]{\href{#1}{ADS}}

\title[Optimising Spectroscopic \& Photometric Galaxy Surveys: Same-sky Benefits for Dark Energy \& Modified Gravity]
{Optimising Spectroscopic and Photometric Galaxy Surveys: Same-sky Benefits for Dark Energy and Modified Gravity}
\author[Donnacha Kirk, Ofer Lahav, Sarah Bridle, Stephanie Jouvel, Filipe B. Abdalla, Joshua A. Frieman]{Donnacha Kirk$^{1}$, Ofer Lahav$^{1}$, Sarah Bridle$^{2}$, Stephanie Jouvel$^{3}$, \newauthor Filipe B. Abdalla$^{1}$, Joshua A. Frieman$^{4}$\\
$^{1}$Department of Physics \& Astronomy, University College London, Gower Street, London, WC1E 6BT, UK\\
$^{2}$Jodrell Bank Centre for Astrophysics, School of Physics and Astronomy, The University of Manchester, Manchester, M13 9PL, UK\\
$^{3}$Institut de Ci\`{e}ncies de l'Espai (ICE, IEEC/CSIC), E-08193 Bellaterra (Barcelona), Spain\\
$^{4}$Fermilab Center for Particle Astrophysics, Batavia, IL 60510 \\
$^{4}$Kavli Institute for Cosmological Physics, The University of Chicago, Chicago, IL 60637}

\begin {document}

\pagerange{\pageref{firstpage}--\pageref{lastpage}} \pubyear{2009}

\maketitle

\label{firstpage}

\begin{abstract}
The combination of multiple cosmological probes can produce measurements of cosmological parameters much more stringent than those possible with any individual probe. We examine the combination of two highly correlated probes of late-time structure growth: (i) weak gravitational lensing from a survey with photometric redshifts and (ii) galaxy clustering and redshift space distortions from a survey with spectroscopic redshifts. We choose generic survey designs so that our results are applicable to a range of current and future photometric redshift (e.g. KiDS, DES, HSC, Euclid) and spectroscopic redshift (e.g. DESI, 4MOST, Sumire) surveys. Combining the surveys greatly improves their power to measure both dark energy and modified gravity. An independent, non-overlapping combination sees a dark energy figure of merit more than 4 times larger than that produced by either survey alone. The powerful synergies between the surveys are strongest for modified gravity, where their constraints are orthogonal, producing a non-overlapping joint figure of merit nearly 2 orders of magnitude larger than either alone. Our projected angular power spectrum formalism makes it easy to model the cross-correlation observable when the surveys overlap on the sky, producing a joint data vector and full covariance matrix. We calculate a same-sky improvement factor, from the inclusion of these cross-correlations, relative to non-overlapping surveys. We find nearly a factor of 4 for dark energy and more than a factor of 2 for modified gravity. The exact forecast figures of merit and same-sky benefits can be radically affected by a range of forecasts assumption, which we explore methodically in a sensitivity analysis. We show that that our fiducial assumptions produce robust results which give a good average picture of the science return from combining photometric and spectroscopic surveys.
\end{abstract}


\section{Introduction}

The era of ``precision cosmology'' is now a reality. Different cosmological probes are able to measure some of the most fundamental properties of our Universe from the Cosmic Microwave Background (CMB) \citep{planck1,spt,act} to the type Ia supernovae (SNe) \citep{riessea98,perlmutterea99} 
which chart the accelerating expansion of the Universe. Large volume surveys of galaxies and galaxy clusters \citep{Eisenstein_sdss,colless_2df_2003} chronicle the growth of cosmic structure and Weak Gravitational Lensing (WGL) \citep{hoekstra_jain_2008,cfhtlens} gives us, through the bending of light, access to dark matter, the dominant matter species, invisible to direct observation.

The next decade will bring an even greater wave of data as many of these cosmic probes are scaled up to cover more area on the sky, greater volumes and more objects. We detail a number of these surveys in tables \ref{tab:specz_surveys} and \ref{tab:photoz_surveys} \citep{DES_forecasts,Euclid_reference,DESpec_whitepaper,4MOST_whitepaper,BigBOSS_whitepaper,sumire_whitepaper}. Each probe of cosmology requires an enormous effort to understand both the underlying physics and subtle systematic and observational effects as well as the creation of innovative new statistical techniques to deal with the sheer quantity of data being produced. In engineering terms these projects are often pushing boundaries in terms of space science, optics, detector design, computation and data storage. Cosmological probes are generally complementary, in that each probes a different combination of the cosmological parameters we are interested in, while being sensitive to different sets of nuisance parameters and systematics.

While each different cosmological probe will gather data of unprecedented precision over the next decade and beyond, it is already clear that the strongest constraints on cosmology come from the proper combination of different probes \citep{kilbinger_cfhtlens_2013,jee_combinedprobes_2013}. These combinations break degeneracies between cosmological (and nuisance) parameters and allow a level of precision much beyond any individual probe. Indeed this is the source of our current ``concordance cosmology'', $\Lambda$CDM \citep{komatsu_wmap7_2011}. Some cosmological probes are relatively independent, perhaps the CMB and SNe are a good example. These probes can be combined in a very simple way without worrying about the cross-talk between observables or double counting of information. This, however, is the exception. Most probes are highly correlated as they probe the same underlying physical processes, whether that is the expansion history of the Universe or the perturbations of the large-scale gravitational potential as it evolves with time. 

Given this situation, increasing attention is being paid to the correct way to combine multiple cosmological probes. While the relatively independent probes we mentioned can be treated separately and combined on the level of multiplied posterior probabilities, this is not possible with the late-time Large-Scale Structure (LSS) probes which are highly correlated both in terms of cosmological information and in systematic effects. In these cases it is essential to construct a joint data vector which can model all cosmological and systematic effects simultaneously, including their cross-correlations, and avoid double counting. In addition one should perform a simultaneous joint likelihood analysis using a covariance matrix which includes all cross correlation terms (and off-diagonal elements) between the different probes. If these complications are ignored the final result can be strongly biased \citep{eifler_2013_covariance,taylor_joachimi_kitching_2013}.

For clarity this paper concentrates on the combination of two types of survey which will become available over the next 5-10 years. We choose a large area optical cosmic shear survey with photometric-quality redshifts, modelled on the Dark Energy Survey (DES) (5000 deg$^2$ with $\sim$200 million galaxies) and a medium scale spectroscopic LSS survey (5000 deg$^2$ targeting $\sim$10 million galaxies) similar to the DESI (combined Big-BOSS, DESpec), 4MOST and Sumire concepts \citep{DESpec_whitepaper,4MOST_whitepaper,BigBOSS_whitepaper,sumire_whitepaper}. See tables \ref{tab:specz_surveys} and \ref{tab:photoz_surveys} for more details on current and future surveys. Although there are many possible analyses one can make with the wealth of data provided by these two types of survey (DES alone combines information from WGL, LSS, galaxy clusters and SNe) we choose to limit ourselves to WGL from the photometric redshift (photo-z) survey and LSS (galaxy power spectrum including RSDs) from the spectroscopic redshift (spec-z) survey. This pared down approach allows us to explore the impact of nuisance parameter modeling \& choice, survey strategy and survey overlap in a clean way without having to deal with too many competing effects. For the same reason we choose to model both probes and their cross-correlations in the same projected angular power spectrum formalism.

The combination of a photo-z WGL survey and a spec-z galaxy clustering survey has been studied by a number of papers including \citet{cai_bernstein_2012,gaztanaga_pau_2012,duncanea_2013,deputter_dore_das_2013}. In general these papers have modelled different observables using different formalisms. Our approach in this paper is to model both WGL and galaxy clustering, including RSDs as projected angular power spectra, $C(l)$ \citep{hu99,bernstein_2008}. While there may be some loss in accuracy for the spec-z case due to projection along redshift we are interested in presenting a unified framework in which each observable is treated on the same footing and cross-correlations can be handled naturally. This fits with the philosophy of jointly modelling all cosmological/systematic effects in the same `combined probes' data vector and a single joint covariance matrix. In the same spirit we try to make explicit all assumptions about observable/survey modelling and the treatment of nuisance parameters. For the most fundamental assumptions we examine the impact of varying each independently as a sensitivity analysis. A full ``optimisation'' would vary these assumptions simultaneously and search for the best combination but we think many are currently so ill-understood that it is more important to disentangle the separate effects. Each assumption will require specialist attention to settle on a ``correct'' approach, we hope merely to demonstrate the power of these assumptions to change survey results and the need for detailed further attention. 

This paper forms a companion piece to \citet{jouvelea_2013}. We model similar surveys but, as a division of labour, we restrict consideration of target selection, survey design and observing strategy to \citet{jouvelea_2013}. This paper considers assumptions on theoretical formalism, systematics including galaxy bias \& photo-z error, survey overlap and more. Assumptions varied in \citet{jouvelea_2013} are fixed in this paper and vice versa.

Section \ref{sec:survey_landscape} talks about the landscape of photo-z and spec-z surveys. Then in section \ref{sec:unified_cls} we present our $C(l)$s formalism for both cosmic shear and galaxy clustering before detailing the rest of our assumptions about nuisance parameters and fiducial survey strategies in section \ref{sec:forecasting_assumptions}. Our forecast constraints on DE and MG are given in section \ref{sec:forecasts}, where each subsection details the impact of a move away from our fiducial assumptions. We draw together the implications of these results in section \ref{sec:discussion} before concluding in section \ref{sec:conclusions}. 

\section{Photometry \& Spectroscopy}
\label{sec:survey_landscape}

When we make a survey of galaxies in the Universe, whether to study Redshift Space Distortions (RSDs), Weak Gravitational Lensing (WGL), Baryon Acoustic Oscillations (BAOs) or galaxy clustering itself, we need to characterise the position of each galaxy using three coordinates. Two of these (commonly RA \& DEC) locate the galaxy in two dimensions on the plane of the sky. It is relatively straightforward to achieve a precise measurement of sky position, with accuracies of sub-arcsecond achievable even for ground based observations. In contrast, fixing the third coordinate, the galaxy's distance from the observer along the line of sight, is considerably more challending. We measure a galaxy's redshift, the lengthening of the wavelength of light from that galaxy as it recedes from us under the Hubble flow, and use it to determine distance. More distant objects have greater redshifts. In general a measurement of distance requires assumptions to be made about cosmological parameters while a redshift measurement does not. 

The most accurate method for determining a galaxy's redshift is spectroscopy. Light from the galaxy is split into its frequency components and the movement of spectral features to the red is used to measure an accurate redshift. Spectroscopic redshifts (spec-z) can reach an accuracy of better than $10^{-3}$, however this process is costly and time consuming. Each galaxy must be examined individually and observed for sufficient time that enough light is collected and a clear spectrum established. Modern multi-object spectrographs expedite this process by using multiple optical fibres to collect light for up to 4,000 galaxies simulataneously. However even these cutting edge, high-throughput machines are limited to observing $\sim$60,000 galaxy spectra per observing night \citep{DESpec_whitepaper}.

There is a faster but less accurate redshift estimation technique in common use for large optical surveys. Known as photometric redshift (photo-z) estimation, it dispenses with the spectrograph entirely and relies on the fact that a standard optical survey will observe in multiple frequency bands (u, g, r, i, z, y etc.), recording images for each exposure under each filter on a many mega-pixel CCD camera. Combining intensity information for a single object from multiple filters produces what is in effect a very low resolution galaxy spectrum which can be used to estimate redshift. These techniques are limited to an accuracy of $\sigma_z = \delta_z (1+z) \approx 0.07 (1+z)$ for a ground-based survey or $\sigma_z \approx 0.05 (1+z)$ for a space-based survey each using $\sim 5$ filter bands. The benefit is that they are significantly faster than equivalent spec-z surveys, capturing a couple of orders of magnitude more galaxies per observing night \citep{hildebrandt_2012_cfhtlens}. 

A new generation of high-resolution photometric surveys, such as PAU \citep{Benitez_PAU}, are also planned. These aim to fill a gap between the standard spec-z and photo-z surveys by using up to 50 filter bands to achieve photometric redshift reconstructions of much greater accuracy. 
Surveys of all these types are major investments in terms of money, instrument time and staff-time with observing time alone counted in hundreds of nights.

In general, the quality of imaging surveys is essential for the success of a multi-object spectroscopic survey on different levels, 
at increasing demand on image quality: (i) imaging is critically required to create a catalogues of objects for fibre allocation;
(ii) the photometric quality and number of filters impact the success rate of selection of LRGs,
ELGs, and $z > 2.1$ QSO’s; (iii) the images can be used for shape measurements for weak lensing
(cosmic shear) and hence enhance the science as described below; (iv) many other science byproducts
may result from combining imaging and spectroscopy, e.g. for detailed studies of
galaxy evolution. Combining imaging and spectroscopy could be useful for cross-calibration
techniques for photo-z testing \citep{zhang09}, and the cross-correlation between surveys to provide clustering
measurements that are robust to systematics \citep{yoo_seljak_2012}. Detailed discussions on these issues are given in
both the BigBOSS and DESpec white papers \citep{DESpec_whitepaper,BigBOSS_whitepaper}, and in the report of the Joint Working Group
BigBOSS-DES\footnote{http://www.astronomy.ohio-state.edu/~dhw/jwg.pdf}. 

There are a number of suitable photo-z and spec-z surveys already available with many more in progress or due to start over the coming years on different parts of the sky. As results from more surveys become available, the optimal use of overlapping photo-z/spec-z sky area will become a crucial question if we are to obtain the best constraints on cosmology from the available data.  We summarise some current and future spectroscopic surveys in table \ref{tab:specz_surveys} and do the same for photometric surveys in table \ref{tab:photoz_surveys}.

It has been illustrated in a number of papers \citep{bernstein_cai_2011,cai_bernstein_2012,gaztanaga_pau_2012,DESpec_whitepaper,BigBOSS_whitepaper,duncanea_2013} that a combination of Redshift
Space Distortion (RSD) from spectroscopic surveys and
weak lensing from imaging surveys is a very powerful tool to constrain Dark Energy and
deviations from General Relativity. 
Weak lensing and BAO/RSD are unique probes of large-scale structure, exploring different
scales in k-space, where a combined analysis may be able to remove some underlying
degeneracies. 

Furthermore, having both the spectroscopy and imaging on the same part of sky
could provide access to new cosmological tests. When observed on the same part of the sky, the
galaxies observed with a spectroscopic survey map the underlying mass fluctuations that lead in turn to the weak lensing of
the distant imaged galaxies. This constrains directly galaxy biasing \citep{gaztanaga_pau_2012}, reduces
cosmic variance \citep{mcdonald_seljak_2009}, and it improves photo-z determinations \citep{newman_2008_photoz,zhang09}
Our calculations below and \citep{gaztanaga_pau_2012} show that DESI-like  spectroscopic redshifts combined
with a high quality imaging survey boosts our ability to measure the DETF Figure of Merit. This
improvement is stronger when the spectroscopy and imaging overlap . Tests of General Relativity benefit even more from the combinations of RSD and weak
lensing because each responds differently to combinations of the two metric potentials. Again,
having same sky configuration gains an additional improvement. However, we note that other
calculations (e.g. BigBOSS White Paper, \citet{cai_bernstein_2012,deputter_dore_das_2013}) do not find improvement from “same sky”.
The source of the discrepancy maybe due to the implementation of the covariance matrix
calculations, assumptions about galaxy biasing for LSS and intrinsic alignment for WGL, the
range of k-values, and the assumed sky area and the redshift distribution of the spectroscopic
sample. 

\begin{table*}
   \centering
   \begin{tabular}{ |l|l|c|l|c|c| } 
   \hline
Instrument & Telescope & No. Galaxies & Sq. Deg. \\
\hline
SDSS I + II & APO 2.5m    & 85K LRG 					& 7,600  \\
\hline
Wiggle-Z    & AAT 3.9m    & 239K 						& 1,000  \\
\hline
BOSS 		& APO 2.5m    & 1.4M LRG + 160K Ly-$\alpha$ & 10,000 \\
\hline
HETDEX      & HET 9.2m    & 1M 						    & 420 	 \\
\hline
eBOSS 		& APO 2.5m    & 600K LRG + 70K Ly-$\alpha$  & 7,000  \\
\hline
DESI 	& NOAO 4m     & 32M LRG + 2M Ly-$\alpha$    & 18,000 \\
\hline
SUMIRE PFS  & Subaru 8.2m & 4M 					 	    & 1,400  \\
\hline
4MOST 	    & VISTA 4.1m  & 6-20M bright objects 	    & 15,000 \\
\hline
EUCLID 	    & 1.2m space  & 75M 				 	    & 14,700 \\
\hline
   \end{tabular}
   \caption{Summary of current or planned BAO capable spectroscopic surveys. Based on table 4 of the MS-DESI Science Alternatives Report. \citep{wigglez,BOSS,eBOSS,hetdex,sumire_pfs,4MOST_whitepaper,Euclid_reference,DESpec_whitepaper,BigBOSS_whitepaper},http://www.sdss.org}
   \label{tab:specz_surveys}
\end{table*}

\begin{table*}
   \centering
   \begin{tabular}{ |l|l|c|c|l|c|c| } 
   \hline
Instrument & Telescope & Observing Bands & No. Galaxies & Sq. Deg. \\
\hline
DES & Blanco 4m & $g, r, i, z, y$ & 300M & 5000 \\
\hline
KiDS & VST 2.6m & $u, g, r, i$ & 90M & 1500 \\
\hline
VHS & Vista 4m & $Y, J, H, Ks$ & 400M & 20,000 \\
\hline
Viking & Vista 4m & $Z, Y, J, H, K$ & - & 1,500 \\
\hline
HSC & Subaru 8.2m & $g, r, i, z, y$ & 400M & 2,000 \\
\hline
Pan-STARRS 1 & Hawaii 1.8m & $g, r, i, z, y$ & 1B & 30,000 \\
\hline
PAU & WHT 4m & 40 narrow-band & 30,000 & 100-200 \\
\hline
J-PAS & OAJ 2.5m & 54 narrow-band & 14M LRG & 8,000 \\
\hline
Skymapper & SSO 1.35m & $u, v, g, r, i, z$ & - & 20,000 \\
\hline
LSST & LSST 8.4m & $u, g, r, i, z, y$ & 4B & 20,000 \\
\hline
Euclid & 1.2m space & $R+I+Z, Y, J, H$ & 1.5B	 & 14,700 \\
\hline
   \end{tabular}
   \caption{Summary of current or planned photometric surveys for LSS and/or WGL. \citep{DES_forecasts,kids,viking,pau,jpas,skymapper,Euclid_reference,lsst} http://www.ast.cam.ac.uk/~rgm/vhs/, http://www.naoj.org/Projects/HSC/, http://ps1sc.org/ }
   \label{tab:photoz_surveys}
\end{table*}

\section{A unified $C(l)$s framework}
\label{sec:unified_cls}

In this paper we have made a decision to describe all our observables, for both the spec-z and photo-z surveys, using projected angular power spectra, $C(l)$s. This enables us to use the same formalism for cosmic shear, LSS and RSDs. More importantly it provides a language in which the cross-correlation between probes, their joint data vector and joint covariance can be written without resort to any special machinery or complicated, untested derivations. In addition we can include systematic effects in a consistent way for all probes. This paper treats galaxy bias, the galaxy-shear correlation coefficient and photometric redshift error. It is straightforward to expand to Intrinsic Alignments (IAs) \citep{joachimi_bridle_2009} and other systematics.

A general $C(l)$ is the projection of two window functions where each corresponds to the projection kernel of a particular observable for a particular tomographic bin. The projected angular power spectrum for observable X in bin $i$ and observable Y in bin $j$ is given by
\begin{equation}
C_{XY}^{i,j}(l) = \frac{2}{\pi} \int W_{X}^{i}(l,k)W_{Y}^{j}(l,k) k^{2} P(k) \textrm{d}k,
\label{eqn:Cls}
\end{equation}
where $W_{X}^{i}(l,k)$ is the window function for observable X, tomographic bin $i$. We describe the window functions used in this paper in section \ref{sec:weight_fns} below. $P(k)$ is the nonlinear matter power spectrum today, $k$ denotes wavenumber, measured in $hMpc^{-1}$, and $l$ denotes angular multipole \citep{fishersl94}. 

We consider cosmic shear, denoted by $\epsilon$, and galaxy clustering, denoted by $n$. This gives us three different $C(l)$ observables: $C_{nn}^{ij}(l)$, the galaxy-galaxy correlation (elsewhere called galaxy clustering), $C_{\epsilon\epsilon}^{ij}(l)$, the shear-shear correlation from WGL and $C_{n\epsilon}^{ij}(l)$, the galaxy-shear cross-correlation. 

We assume that we have access to two surveys: an optical survey of ~300 million galaxies with photometric quality redshifts and sufficient resolution to perform shape measurement for WGL, and a survey of ~10 million galaxies with spectroscopic quality redshifts. We will make use of these surveys throughout the paper and refer to them as our ``photo-z survey'' and our ``spec-z survey''. We split each survey into a number of tomographic bins in redshift. The spec-z quality redshift allows us to bin these galaxies at much higher resolution in z. We choose 5 tomographic bins for the photo-z survey (consistent with DES-like surveys in the literature) and 40 tomographic bins for our spec-z survey (still computationally feasible and giving sufficient redshift resolution to capture the bulk of the available information \citep{asoreyea_2012}). For simplicity we assume our cosmic shear observable, $\epsilon$ is always from the photo-z survey while our galaxy clustering observable, $n$, comes from the spec-z survey. The cross-correlation observable $n\epsilon$, where it is present, uses galaxies from both surveys. More details on our fiducial survey assumptions are given in section \ref{sec:forecasting_assumptions}.

\subsection{Weight Functions}
\label{sec:weight_fns}

The projected angular power spectrum for a particular probe and combination of tomographic bins is obtained by including the appropriate weight functions in the general $C(l)$s equation, given in eqn. \ref{eqn:Cls}, above. The probes we consider are cosmic shear, $\epsilon$, and galaxy clustering, $n$.

The weight function for galaxy clustering is
\begin{equation}
W_{n}^{i}(l,k) = \int b_{g}(k,z)n^{i}(z)j_{l}(k\chi(z)) D(z) dz,
\label{eqn:W_n}
\end{equation}
where $b_{g}(k,z)$ is the galaxy bias, $n^{i}(z)$ is the galaxy redshift distribution of tomographic bin $i$, $D(z)$ is the linear growth function and $j_{l}(k\chi(z))$ is the $l$-th order speherical Bessel function of the first type \citep{huterer_2001}.

The weight function for cosmic shear is 
\begin{equation}
W_{\epsilon}^{i}(l,k) = \int q^{i}(z)j_{l}(k\chi(z)) D(z) dz,
\label{eqn:W_G}
\end{equation}
where $q^{i}(z)$ is the lensing weight function, given by
\begin{equation}
q^{i}(z) = \frac{3H_{0}^{2}\Omega_m}{2c^2}\frac{\chi(z)}{a(z)}\int_{\chi_{hor}}^{\chi}d\chi' n^{i}\left(\chi(z')\right)\frac{\chi(z')-\chi(z)}{\chi(z')},
\end{equation}
where $\chi$ is comoving distance, $n^{i}(\chi)$ is the galaxy redshift distribution of tomographic bin $i$ \citep{takadaj04_cospars,joachimi_bridle_2009}.

Each of these weight functions are constructed for a particular tomographic redshift bin, $i$, defined by the galaxy redshift distribution, $n^{i}(z)$. Together we can use these weight functions to define three probes based on 2-point functions: the shear-shear correlation, $\epsilon\epsilon$, the galaxy-galaxy correlation, $nn$, and the galaxy-shear cross-correlation, $n\epsilon$.

\subsection{Redshift Space Distortions}
\label{sec:RSDs}
When considering a survey with sufficiently high resolution redshift information it is possible to learn more about LSS than the galaxy positions alone provide. Galaxies, as well as moving as part of the underlying Hubble flow, have their own peculiar velocities, sourced by local gravitational potentials, past mergers etc. In a galaxy redshift survey a net peculiar velocity along the line of sight away from (towards) the observer adds to (subtracts from) the apparent redshift of a given galaxy. The impact of these effects on the survey are known as Redshift Space Distortions (RSDs) and can be used to learn about cosmology as they are sourced by the local gravitational potential. 

The distortion caused by coherent infall velovities takes a particularly simple form in Fourier space, given by the familiar Kaiser formula \citep{kaiser_1987_RSDs}
\begin{equation}
\delta^{s}_{g}(k,\mu) = (b_{g}(k,z)+f(z)\mu^{2})\delta_{m}(k)
\end{equation}
where $\mu$ is the cosine of the angle between $k$ and the line-of-sight, the superscript $s$ denotes redshift-space, $b_{g}(k,z)$ is the galaxy bias and and $f(z) = \frac{dlnD(z)}{dlna} \approx \Omega_{m}^{0.55}$ \citep{peebles_book}.  

We extend the galaxy window function (eqn.\ref{eqn:W_n}) to include the effects of RSDs following \citet{fishersl94,heavens_taylor_95,padmanabhanea05} where they express the RSDs as an additional term in the galaxy clustering window function. The total LSS weight function is given by

\begin{equation}
W_{n,tot}^{i}(l,k) = W_{n}^{i}(l,k) + W_{n,R}^{i}(l,k). 
\end{equation}

That is, the sum of the non-RSD weight function (eqn.\ref{eqn:W_n}) and a new RSD term 
\begin{equation}
\begin{split}
W_{n,R}^{i}&(l,k) = \beta\int f(y) \Bigl[  \frac{(2l^{2}+2l-1)}{(2l+3)(2l-1)}j_{l}(ky) \\
& -\frac{l(l-1)}{(2l-1)(2l+1)}j_{l-2}(ky) - \frac{(l+1)(l+2)}{(2l+1)(2l+3)}j_{l+2}(ky) \Bigr] dy.
\end{split}
\end{equation}

This approach does not model the `Finger of God' effect- small scale RSDs due to the virial motion of galaxies within clusters \citep{kang_fingerofgod_2002}. As we cut our galaxy observables to exclude non-linear scales (see section \ref{sec:theory_nl} below) we feel justified in ignoring this effect in our model. The lensing kernel is broad enough to wash out any effects from RSDs so we leave eqn. \ref{eqn:W_G} unchanged and ignore RSDs for our WGL observables.

\section{Forecasting Assumptions}
\label{sec:forecasting_assumptions}

Here we detail the fiducial assumptions we make when forecasting the science results of our photo-z and spec-z surveys. We begin with a summary of the Fisher Matrix (FM) formalism which we use to make our forecasts and the way in which this, combined with our $C(l)$s approach, lends itself readily to the combination of probes from different surveys.

Next we describe the survey strategy and target selection assumptions we make for both surveys. These are generally held fixed in this paper but are explored in detail for the spec-z survey in our companion paper \citep{jouvelea_2013}.

The $C(l)$s formalism described above is general for cosmic shear and galaxy clustering (including RSDs) in a $\Lambda$CDM cosmology. Here we describe in detail the assumptions we make on elements of the formalism including galaxy bias, the galaxy-shear cross-correlation coefficient, non-linear clustering and the range of scales considered. We also describe an extension to the formalism which describes deviations from General Relativity (GR). It is this extension which allows us to forecast the ability of our probes to constrain deviations from GR.  Several of these assumptions are subsequently varied in section \ref{sec:forecasts} where we study their impact on the constraining power of the individual surveys and their joint combinations.

\subsection{Fisher Matrices}
\label{sec:FMs}

We make our forecats under the Fisher Matrix formalism. To forecast constraints on cosmological parameters we calculate the Fisher information matrix (see e.g.\citet{heavens_statistics_2009}), which is the expectation value of the Hessian matrix of the log likelihood with respect to some parameters $\alpha, \beta$,
\begin{equation}
F_{\alpha\beta} \equiv \left< H_{\alpha\beta} \right> = \left< -\frac{\partial^{2}lnL}{\partial\theta_{\alpha}\partial\theta_{\beta}} \right>
\end{equation}
and can be written 
\begin{equation}
F_{\alpha\beta} = \sum^{l_{max}}_{l=l_{min}} \sum_{(i,j),(m,n)} \frac{\partial D^{i,j}(l)}{\partial p_{\alpha} }{\rm Cov}^{-1} \left[ D^{ij}(l),D^{mn}(l) \right] \frac{\partial D^{mn}(l)}{\partial p_{\beta} },
\end{equation}
 where $D(l)$ is the data vector under consideration, $\rm Cov \left[ D_{ij}(l),D_{mn}(l) \right]$ is the covariance matrix, $p_{\alpha}$ label the cosmological and nuisance parameters we vary in our analysis and $i,j$ label pairs of tomographic bins \citep{joachimi_bridle_2009}. The FM formalism provides an estimate of the the marginalised error on each cosmological parameter through the Cramer-Rao inequality,  $\sigma_i \geq \sqrt{(F^{-1})_{\alpha\alpha}}$. This is the marginalised parameter error, the independent error on parameter $\alpha$, i.e. the error if all other parameters are fixed, is given by $\sigma_i \geq \sqrt{1/F_{\alpha\alpha}}$.

\subsection{Combined Probes}
\label{sec:combined_probes}

The above formalism shows how cross-correlations between observables come naturally in the $C(l)$s formalism. In our case we can trivially construct a data vector including both galaxy and shear correlations and their cross-correlations, $D^{i,j}(l) = \{ C_{\epsilon\epsilon}^{ij}(l), C_{n\epsilon}^{ij}(l), C_{nn}^{ij}(l) \}$. This models the available cosmological information when our two surveys overlap on the sky. It is also a simple matter to calculate the full covariance matrix between observables, including all the off-diagonal elements \citep{joachimi_bridle_2009,bernstein_2008},
\begin{equation}
\textrm{Cov}(l) = \left( \begin{tabular}{c|c|c}
$\textrm{Cov}_{\epsilon\epsilon\epsilon\epsilon}^{ijkl}(l)$ & $\textrm{Cov}_{\epsilon\epsilon n\epsilon}^{ijkl}(l)$ & $\textrm{Cov}_{\epsilon\epsilon nn}^{ijkl}(l)$ \\ 
$\textrm{Cov}_{n\epsilon\epsilon\epsilon}^{ijkl}(l)$ & $\textrm{Cov}_{n\epsilon n\epsilon}^{ijkl}(l)$  & $\textrm{Cov}_{n\epsilon nn}^{ijkl}(l)$ \\ 
$\textrm{Cov}_{nn\epsilon\epsilon}^{ijkl}(l)$ & $\textrm{Cov}_{nnn\epsilon}^{ijkl}(l)$ & $\textrm{Cov}_{nnnn}^{ijkl}(l)$ \\
 \end{tabular} \right)
\end{equation}
We calculate each individual covariance sub-matrix as
\begin{equation}
\begin{split}
\textrm{Cov}^{(ijkl)}_{\alpha\beta\gamma\delta}(l) &\equiv \left<\Delta C_{\alpha\beta}^{(ij)}(l) \Delta C_{\gamma\delta}^{(kl)}(l') \right> \\
&= \delta_{ll'}\frac{2\pi}{Al\Delta l}\left\{ \bar{C}^{(ik)}_{\alpha\gamma}(l)\bar{C}^{(jl)}_{\beta\delta}(l) +\bar{C}^{(il)}_{\alpha\delta}(l)\bar{C}^{(jk)}_{\beta\gamma}(l) \right\} 
\end{split}
\end{equation}
where $\bar{C}$ accounts for shot and shape noise as:
\begin{equation}
\bar{C}_{\alpha\beta}^{(ij)}(l) \equiv C_{\alpha\beta}^{(ij)}(l) + N_{\alpha\beta}^{(ij)}
\end{equation}
and the noise contributions are given by
\begin{eqnarray}
N_{\epsilon\epsilon}^{(ij)} &=& \delta_{ij}\frac{\sigma_{\epsilon}^{2}}{2\bar{n}_{g}^{(i)}} \\
N_{nn}^{(ij)} &=& \delta_{ij}\frac{1}{\bar{n}_{g}^{(i)}} \\ 
N_{n\epsilon}^{(ij)} &=& 0 
\end{eqnarray}
where $\sigma_{\epsilon}^{2}$ is the shape noise from cosmic shear galaxy shape measurement and $\bar{n}_{g}$ is the shot noise due to the fact we observe a finite number of galaxies. 

If the only observables being considered are $\epsilon\epsilon$ or $nn$ then only the covariance sub-matrices ${Cov}_{\epsilon\epsilon\epsilon\epsilon}^{ijkl}(l)$ and ${Cov}_{nnnn}^{ijkl}(l)$ respectively need be considered.

We recognise that the projected angular power spectra formalism has some limitations. Even with a large number of tomographic bins, there is still a loss of information due to projection along the line of sight within bins which, after all, have some finite width. While several analyses have shown that this formalism obtains all the available cosmological information from a photo-z WGL survey \citep{joachimi_bridle_2009,MGPaper2}, it is likely that some information is lost in our analysis of the spec-z LSS survey compared to a full 3D analysis. Nevertheless, we feel justified in using a $C(l)$s approach as \citet{asoreyea_2012} have shown it to be competitive even for spec-z surveys and it provides many benefits in terms of joint systematic and covariance matrix estimation not available to a mixed $C(l)$/$P(k)$ approach.

\subsection{Cosmological Parameters}
\label{sec:cospars}

We assume a fiducial set of late-Universe cosmological parameters, $p_{\alpha} = \{ \Omega_m,w_0,w_a,h,\sigma_8,\Omega_b,n_s\} = \{ 0.25, -1, 0, 0.7, 0.8, 0.05, 1 \}$, where $\Omega_m$ \& $\Omega_b$ are the dimensionless matter and baryon densities respectively (i.e. $\Omega_{cdm} = \Omega_{m} - \Omega_{b}$), $w_0$ and $w_a$ parameterise the DE equation of state, $w(z) = w_0 + w_{a}z/(1+z)$, $h$ is the Hubble parameter, $\sigma_8$ is the normalisation of the matter power spectrum, $n_s$ is the slope of the primordial power spectrum and $\delta_z$ is the error on the photometric galaxy redshift distribution. We also use a number of nuisance parameters for galaxy bias, $b_X$, described in section \ref{sec:bg} below. We allow a single global photo-z error nuisance parameter, $\delta_z$, described in section \ref{sec:theory_photoz}. All quoted results are marginalised over this parameter space. We assume flatness throughout as a theoretical prior. We apply wide, flat, uninformative priors to all cosmological and nuisance parameters. We also, where appropriate, assume two modified gravity (MG) parameters $Q_0$ \& $R_0$, a set of nuisance parameters for the galaxy-shear cross-correlation coefficient, $r_g$, and an extended set of photo-z nuisance parameters. These are detailed in sections \ref{sec:MG}, \ref{sec:bg_rg} and \ref{sec:photoz_error} below.

\subsection{Figures of Merit}
\label{sec:FoMs}

It is often convenient to summarise the cosmological constraining power of a survey or combination of surveys in one number. This is of course a great simplification which pays no attention to many potential benefits of a particular survey design but it has the benefit of allowing easy comparison between survey designs, assumptions and even the results of different papers.

Throughout this paper we will quote values for the Dark Energy Figure of Merit (DE FoM), based on the Dark Energy Task Force \citep{detf} definition,
\begin{equation}
\textrm{FoM}_{\textrm{DE}} = \frac{1}{4\sqrt{\textrm{det}(F^{-1})_{\textrm{DE}}}},
\end{equation}
where the subscript $\textrm{DE}$ denotes the $2 \times 2$ sub-matrix of the inverse FM that corresponds to the entries for the equation of state of DE parameters, $w_0$ and $w_a$. Note that different prefactors to this equation exist in the literature. We use the factor of $1/4$ for consistency with related papers \citep{bridleandking,joachimi_bridle_2009}.

By analogy we define a Modified Gravity Figure of Merit (MG FoM),
\begin{equation}
\textrm{FoM}_{\textrm{MG}} = \frac{1}{4\sqrt{\textrm{det}(F^{-1})_{\textrm{MG}}}},
\end{equation}
where the subscript $\textrm{MG}$ denotes the $2 \times 2$ sub-matrix of the inverse FM that corresponds to the entries for the equation of state of DE parameters, $Q_0$ and $R_{0}\frac{Q_{0}(1+R_{0})}{2}$ \citep{MGPaper2}. 
Our MG parameters are held fixed when the DE FoM is calculated but the DE parameters are allowed to vary when we calculate the MG FoM because our survey combination needs to be able to constrain expansion history as well as deviations from GR (from mis-matched expansion/growth of structure).

\subsection{Survey Strategy}
\label{sec:survey_strategy}

We assume two fiducial surveys: a photo-z WGL survey and a spec-z LSS survey including RSDs. Each is modelled generically so that our results are as widely applicable as possible. Particularly in the spec-z case our toy $n(z)$ does not look like the true redshift distribution that any particular survey would measure but it has the benefits of simplicity, clarity and generality. We can examine the impact of forecast assumptions without dealing with the complicated interplay between z-coverage, nuisance parameters, ell-cuts etc. and a particular, feature-full $n(z)$. We examine the impact of a specific survey target selection in section \ref{sec:steph_nz} and the whole issue is investigated in detail as part of our companion paper \citet{jouvelea_2013}.

We model our photo-z survey on the Dark Energy Survey (DES) which saw first light in September 2012 and is due to start full survey operations in September 2013. We assume 300 million galaxies are observed over 5000 deg$^2$ with a gaussian photometric redshift error of $\delta_z = 0.07(1+z)$. The survey covers a redshift range of $0 < z < 3$ with galaxy redshift distribution assumed to be Smail-type \citep{efstathiou_smail_nz_1991},
\begin{equation}
n(z) =  z^{\alpha} exp \left[ -(\frac{z}{z_{0}})^{\beta}\right]
\end{equation}
with $\alpha = 2$, $\beta = 1.5$ and $z_{0} = 0.8/\sqrt{2}$. The $n(z)$ is split into five tomographic bins of roughly equal number density. We assume a lensing shape noise of $\sigma_{\gamma} = 0.23$. These survey assumptions for the photo-z survey are fixed for all the results presented below.

The spec-z LSS survey we choose to model is representative of a number of near-term spectroscopic surveys, particularly DESI (combined DESpec and BigBOSS), 4MOST and Sumire. We treat the fiducial survey as a central point in the parameter space of survey assumptions, our results below investigate the impact of changing the survey properties one assumption at a time. 
For the fiducial survey we assume 10 million galaxies over 5000 deg$^2$ over redshift $0 < z < 1.7$. We choose $z_{max}=1.7$ as most redshifts in the high-z end, above $z=1$, will be obtained through the redshifted OII line. This line will only be measurable out to a $z_{max}$ which will be roughly 1.7 if our spectrographs lose significant sensitivity at around 1 micron, which is the case with current optical CCDs.  As a toy model we assume a constant number density over this range because we do not want our results to depend on a particular target selection strategy. We assume high quality redshift information, $\delta_z = 0.001(1+z)$, which we use to split the survey into 40 tomographic bins. See our companion paper \citet{jouvelea_2013} for a detailed analysis of how target selection and survey design effects this spec-z survey. In this paper we explore systematic effects and theoretical assumptions. 

Our fiducial survey assumptions are summarised in table \ref{tab:fiducial_surveys}. in this paper we study the impact of survey area and photo-z/spec-z overlap on constraining power but leave other properties fixed for clarity. 

\begin{table}
   \centering
   \begin{tabular}{@{} l|c|c @{}} 
   \hline
Parameter & Photo-z WGL & Spec-z LSS\\
\hline
Area [deg$^2$] & 5000 & 5000\\
$z_{min}$ & 0 & 0\\
$z_{maz}$ & 3 & 1.7\\
$N_z$ & 5 & 40\\
$\delta(z)$ & $0.07(1+z)$ & $0.001(1+z)$\\
$n_g$ [arcmin$^{-2}$] & 10 & 0.56\\
\hline
   \end{tabular}
   \caption{Fiducial survey assumptions for our photo-z WGL survey and spec-z LSS survey. $N_z$ is the number of tomographic redshift bins of equal number density. $\delta_z$ is the gaussian redshift error and $n_g$ is the number density of (usable) targets of the sky. More details on fiducial survey assumptions can be found in section \ref{sec:survey_strategy}.}
   \label{tab:fiducial_surveys}
\end{table}

\subsection{Galaxy Bias}
\label{sec:bg}

As well as our standard cosmological parameters we must also consider a number of `nuisance parameters' which describe systematic effects for which we lack a physical model or the uncertainties on our model contribute significantly to the overall error budget of our experiment. In this work the principle systematic effects are the galaxy bias, $b_g$, and the galaxy-shear correlation coefficitent, $r_g$ \citep{guzik_seljak_2001_rg, mandelbaumea_2013}. For clarity we ignore glaxy intrinsic alignments (IAs), a prime astrophysical systematic in WGL \citep{hiratas10_posterratum,kirk_rassat_host_bridle_2012}, but one beyond the scope of this work. 

When we attempt to model galaxy clustering, including RSDs, for a spec-z survey the most important astrophysical systematic we must consider is galaxy bias. Galaxy bias describes the relationship between the observed galaxy power spectrum and the underlying dark matter power spectrum. We assume linear, local galaxy biasing \citep{baldauf_biasreview_2011}, i.e.
\begin{equation}
P_{nn}(k,z) = b^{2}_{g}(k,z)P_{\delta\delta}(k,z). 
\end{equation}
If $b_{g}(k,z)=1$ then galaxies can be taken as exact, unbiased tracers of the underlying dark matter density. We know this is not the case \citep{jullo_bias_2012}. Galaxies preferentially form in high density environments, making them biased tracers of the dark matter density. In general the relationship between galaxy and dark matter clustering will evolve as a function of redshift and scale. It may also take different forms for different galaxy types. See \citet{swansonea_2010} for a review of a number of current $b_g$ models.

We discuss some physically motivated models for $b_g$ in section \ref{sec:bg_models} below. For now we present our fiducial model which is designed to be agnostic about the form of $b_g$ but to include enough uncertainty in both scale and redshift dependence that we do not over-constrain cosmology due to naive assumptions about our knowledge of bias. Our model follows that of \citet{joachimi_bridle_2009}.

This fiducial model takes the form
\begin{equation}
b_{g}(k,z) = A_{b_{g}}Q_{b_{g}}(k,z),
\end{equation}
where $A_{b_{g}}$ is a variable amplitude parameter with fiducial value $A_{b_{g}}=1$ and $Q_{b_{g}}(k,z)$ is a free function in $k,z$ formed by the interpolation of a grid of $N_{z} \times N_{k}$ nodes in $k,z$, each of which is allowed to vary independently and has a fiducial value of unity. The $N_z$ nodes are spaced linearly throughout the z-range of our survey while the $N_k$ nodes are log spaced between $k = 0.001$ and $30 h Mpc^{-1}$. In total this formalism introduces $1 + N_{z} \times N_{k}$ nuisance parameters for galaxy bias. Our fiducial model assumes a $2 \times 2$ grid for galaxy bias, i.e. 5 nuisance parameters. We investigate the impact of this choice in section \ref{sec:bg_nknz} below.

When we consider the galaxy-shear cross correlation, $C_{n\epsilon}(l)$, there is another parameter that must be defined. The galaxy-shear correlation coefficient, $r_{g}(k,z)$, which mediates between the galaxy and shear power spectra \citep{guzik_seljak_2001_rg,mandelbaumea_2013},
\begin{equation}
r_{g}(k,z) = \frac{P_{n\epsilon}(k,z)}{\sqrt{P_{\epsilon\epsilon}(k,z)P_{nn}(k,z)}}.
\end{equation}
If there is uncertainty about the form of this parameter it can contribute a substantial systematic error to the modelling of the shear-position, $n\epsilon$, cross-correlation. \citet{gaztanaga_pau_2012} argue that, if analysis is restricted to sufficiently large scales and galaxy bias is assumed to be scale-independent, then it is acceptable to set $r_{g}=1$. For most results below we fix $r_g$ at unity but in section \ref{sec:bg_rg} we consider the impact of this assumption when we compare results for $r_{g}=1$ and $r_{g}$ parameterised as $r_{g}(k,z) = A_{r_{g}}Q_{r_{g}}(k,z)$, similarly to the fiducial galaxy bias parameterisation above. In this case another $1 + N_{z} \times N_{k}$ nuisance parameters are introduced to parameterise the cross-correlation. The aim of introducing extra nuisance parameters is to avoid a systematic biasing of our constraints from poor modelling of $r_g$, at the cost of reduced statistical error.

\subsection{Modified Gravity}
\label{sec:MG}

We also present results forecasting the ability of our surveys to constrain deviations from gravity as described by General Relativity (GR). To describe these Modified Gravity (MG) scenarios, we introduce two new parameters, $Q,R$, following the formalism of \citet{MGPaper1,bean_tangmatitham_2010} This formalism treats perturbations to the metric in the Newtonian gauge,
\begin{equation}
ds^2 = -a^{2}(\tau) \left[ (1+2\Psi) \right]d\tau^{2} + a^{2}(\tau) \left[ (1-2\Phi)a^{2})\right] \delta_{ij}dx^{i}dx^{j}
\end{equation}
$Q$ describes changes to the Poisson equation which describes how the Newtonian potential is sourced by matter, 
\begin{equation}
k^{2}\Phi = -4 \pi G Q a^{a} \rho(a) \delta,
\end{equation}
and can be considered as an effective Newton's constant. $R_0$ parameterises the ratio of the Newtonian \& Curvature metric potentials,
\begin{equation}
R = \frac{\Phi}{\Psi}.
\end{equation}
Both $Q$ and $R$ are unity in GR. In general MG theories both could vary as a function of scale and redshift. In this work we assumed, for simplicity, that both are scale independent (our analysis cuts off at scales much larger than those where a screening mechanism would need to be invoked to preserve solar system GR tests) and follow a simple evolution with redshift, $Q = Q_{0}a^{s}$, $R = R_{0}a^{s}$, with $s=3$. This evolution is motivated by the need to preserve CMB and BBN tests of GR at high redshift, allowing the modification to ``turn on'' at late times in an effort to explain cosmic acceleration without invoking Dark Energy. 

We fix $s=3$ but allow $Q_0$ and $R_0$ to vary from their fiducial GR values. When $Q_0 \neq 1$ or $R_0 \neq 1$ the deviation from GR enters our observables in two ways: through changes to the linear growth factor which effects all our observables $\epsilon\epsilon$, $n\epsilon$ and $nn$ as well as the $\beta$ term in the RSD kernel and through changes to the geometric factor in our cosmic shear kernel, adding a factor of $Q(z)(1+R(z))$ to the $n\epsilon$ observable and a factor of $\left[Q(z)(1+R(z))\right]^2$ to $\epsilon\epsilon$.

\subsection{Photometric Redshift Errors}
\label{sec:theory_photoz}
There has been some interest in the use of combined photo-z and spec-z surveys to ``self-calibrate'' the photometric redshift error of the photo-z survey \citep{newman_2008_photoz}. In our fiducial set-up we assume a single global error for the photo-z $n(z)$, $\sigma_z = \delta_z (1+z)$, where $\delta_z$ is a free parameter allowed to vary around its fiducial value of 0.07. We keep this model simple to ensure that our combined results do not rely too heavily on this photo-z calibration effect.

We investigate the impact of this calibration in more detail in section \ref{sec:photoz_error} below. In that section we explore joint constraints in the case of the photo-z error being accurately known and in the more sophisticated case where we allow a variable $\delta_{z}^{i}$ for each tomographic bin $i$ in the photo-z survey. In addition we allow the mean redshift of the photo-z tomographic bins to vary via parameters $\Delta_{z}^{i}$, called photometric redshift bias for bin $i$, with fiducial values zero. This approach gives us a very flexible photo-z error, we can reduce the uncertainty by applying increasingly tight priors to the photo-z nuisance parameters.
 
\subsection{Non-Linear Matter Clustering}
\label{sec:theory_nl}
As matter collapses under gravity, higher density regions reach a point at which their local overdensity is of order unity. In this regime structure formation no longer continues under the well behaved linear growth equations but continues to collapse in a much more complicated non-linear manner. For our forecasts we produce a linear matter power spectrum using the fitting function of \citet{eisensteinhu97} and model the non-linear growth of structure using the halofit model of \citet{smithea03}. However, it is known that this non-linear fitting is only accurate to $\sim$10\% so we are not overly confident of our ability to forecast non-linear growth or, even more so, the non-linear galaxy bias and RSDs at small scales. 

\begin{figure}
  \begin{flushleft}
    \centering
       \includegraphics[width=3in,height=3in]{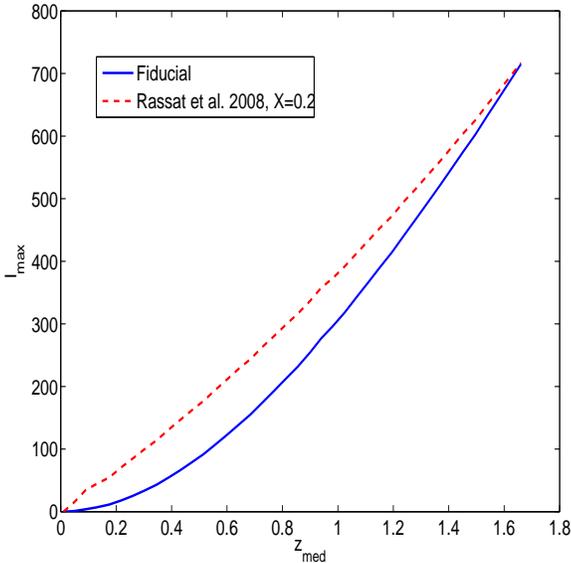}
\vspace{6mm}
\caption{The fiducial $l$-cuts used in this work based on the recipes of \citet{joachimi_bridle_2009} [blue solid] and \citet{Rassatea_2008} with X=0.2 [red dashed]. The plot shows the maximum $l$ value included in the analysis as a function of tomographic bin number, $N_{\textrm{bin}}$. In the case of the $n\epsilon$ correlation, the $n$ bin defines the cut. In the case of the $nn$ correlation we chose the optimistic case and cut on the higher redshift bin. $l_{\textrm{min}}=10$ is assumed throughout. For more details see section \ref{sec:theory_nl}}
\label{fig:lcuts}
  \end{flushleft}
\end{figure}

If naively included the non-linear scales contribute a great deal of constraining power to our LSS observables. Rather than over-constrain cosmology and galaxy bias by including effects which we understand so poorly, we cut non-linear scales from our analysis wherever the $n$ observable appears, i.e. in $C_{nn}(l)$ and $C_{n\epsilon}(l)$. We follow the approach of \citet{Rassatea_2008} by defining some $k_{max}$ which we convert into an $l_{max}(z)$ which defines a maximum multipole per tomographic bin. For consistency we choose $k_{max}=0.132*z^{i}_{med}$ as used in \citet{joachimi_bridle_2009}, where $z^{i}_{med}$ is the median redshift of tomographic bin $i$. It should be noted that this is not exactly the same recipe as the fiducial one set out in \citet{Rassatea_2008}. We show the difference in Fig. \ref{fig:lcuts}. We are happy with our fiducial choice as it is the more conservative of the two. We investigate the impact of these k-cuts based on the \citet{Rassatea_2008} recipe and give more quantitative details in section \ref{sec:kcuts}. Scales smaller than this are removed. Fig. \ref{fig:lcuts} shows the cut in angular scale, $l$, that we deploy as a function of redshift. For the $n\epsilon$ correlation we cut based on the galaxy clustering tomographic bin. The $nn$ correlation offers a choice of tomographic bin to cut on (in the case of cross-bin correlations), we choose the optimistic option and cut on the higher redshift bin. The exact details of this cut are investigated in section \ref{sec:kcuts} below.

\section{Forecasts}
\label{sec:forecasts}

\begin{figure}
  \begin{flushleft}
    \centering
       \includegraphics[width=3in,height=3in]{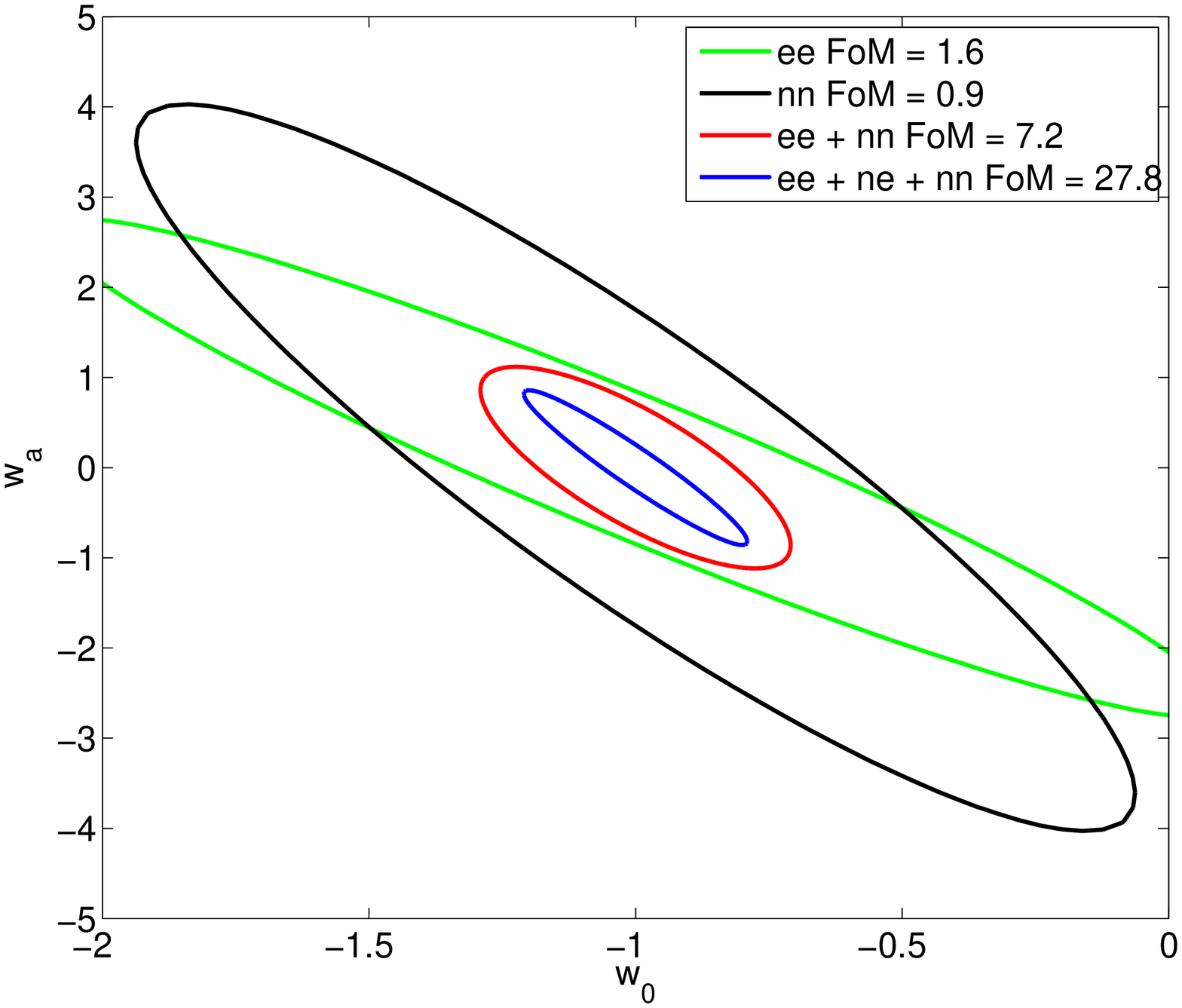}
       \includegraphics[width=3in,height=3in]{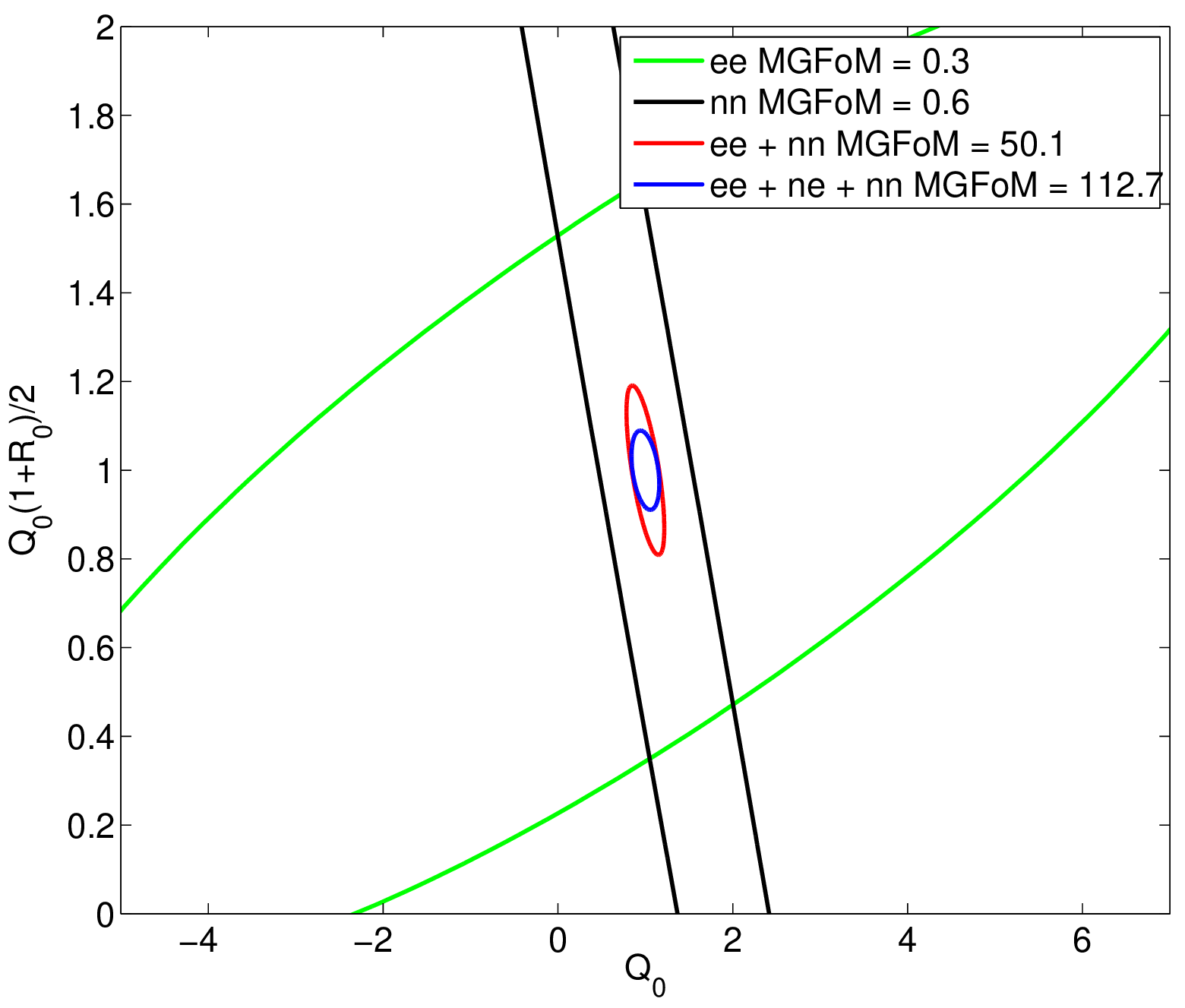}
\vspace{6mm}
\caption{Upper panel: Marginalised $95\%$ errors on $w_0$, $w_a$ for $\epsilon\epsilon$ from our photo-z survey [green contours], $nn$ from our spec-z survey [black contours], their independent combination, $\epsilon\epsilon + nn$, [red contours] \& their combination including cross-correlations, $\epsilon\epsilon+n\epsilon+nn$, [blue contours]. Our fiducial survey strategies are assumed. 5 tomographic bins are used for the photo-z survey, 40 tomographic bins for the spec-z survey. We marginalise over our standard cosmological parameters, assuming GR. We marginalise over galaxy bias, assuming one overall amplitude term and a $2 \times 2$ grid of $b_g(k,z)$ nodes, while fixing $r_g(k,z) = 1$.  We marginalise over a single global photo-z error term. Any observable containing galaxy clustering is assumed to be cut at large ell. RSDs are included for the spec-z survey. Lower panel: Marginalised $95\%$ errors on $Q_0$, $Q_{0}(1+R_{0})/2$ for $\epsilon\epsilon$ from our photo-z survey [green contours], $nn$ from our spec-z survey [black contours], their independent combination, $\epsilon\epsilon + nn$, [red contours] \& their combination including cross-correlations, $\epsilon\epsilon+n\epsilon+nn$, [blue contours]. We marginalise over $w_0$ and $w_a$, all other assumptions are the same as for the upper panel.}
\label{fig:contours_fid}
  \end{flushleft}
\end{figure}

Given the forecasting assumptions detailed in the previous section, we want to answer the primary questions: how well can our example photometric \& spectroscopic surveys constrain cosmology? and what particular benefit do we gain from combining them? Fig \ref{fig:contours_fid} shows $95\%$ confidence contours for DE \& MG respectively, assuming our fiducial survey scenarios. In the case of the DE constraints, GR is assumed. The MG constraints are marginalised over $w_0,w_a$. All other cosmological parameters are marginalised over in both analyses, as are a $2 \times 2$ grid if $b_g$ parameters. $r_g = 1$ is assumed. Both marginalise over our standard set of cosmological parameters, the galaxy bias nuisance parameters and a single photo-z error parameter.

Constraints are shown for the photometric WGL survey alone, $\epsilon\epsilon$, the spectroscopic LSS survey alone , $nn$, the independent combination of the two surveys (this is the constraint from two surveys on separate areas of the sky), $\epsilon\epsilon + nn$, and the dependent combination of both surveys including all cross-correlations (which includes the full information for the case where the survey areas fully overlap), $\epsilon\epsilon + n\epsilon + nn$.

The DE and MG constraints in Fig \ref{fig:contours_fid} share a number of features. In each case the photometric and spectroscopic surveys alone are relatively poorly constraining. This is unsurprising given the cosmology \& nuisance parameters marginalised over and our extremely conservative k-cut to remove non-linear scales where $b_g$ and matter clustering generally are very uncertain. However, even given this limitation, the combination of the spectroscopic $nn$ with the photometric $\epsilon\epsilon$ is extremely beneficial. Adding the surveys ($\epsilon\epsilon+nn$) produces a factor of $\sim 4.5$ improvement in DE FoM compared to the photo-z only survey, without including cross-correlations. When these cross-correlations are included we see a `same-sky benefit' of almost a factor of four in DE FoM. This improvement comes from the inclusion of the $n\epsilon$ cross-correlation observable which introduces new cosmological dependencies and breaks some degeneracies between cosmological parameters and galaxy bias.

In the MG panel of Fig. \ref{fig:contours_fid} the $nn$ only survey is slightly more constraining than $\epsilon\epsilon$ alone. This is due to the ability of the high-resolution RSD measurements to probe variation in the linear growth factor. There is clearly a strong degeneracy between $Q_0$ \& $R_0$ that the $nn$ probe alone cannot overcome.
It requires the extra geometric information contained in the WGL probe to break this degeneracy and produce a closed constraint on deviations from GR. This degeneracy-breaking means that the combination of $\epsilon\epsilon + nn$ produces a much more pronounced improvement than in the DE case, giving more than a factor of 150 improvement over WGL alone. The inclusion of the $n\epsilon$ cross-correlation with its own dependence on $Q_0$, $R_0$ adds another, same-sky, beneift of more than a factor of two. The same-sky benefit is less pronounced for MG than for DE because of the huge improvement gained even from the independent combination of the two probes, making the cross-correlation improvement proportionally less important.

We note that the $nn$ only probe gets nearly all its constraining power for DE and MG from the inclusion of RSDs, without these the marginalistion over $b_g$ reduces the LSS-only constraints to near zero. In this case the full joint constraints rely even more strongly on cross-correlation with the WGL photo-z survey, producing a same-sky benefit factor of $\sim$4.7 and a combined $\epsilon\epsilon + n\epsilon + nn$ DE FoM nearly equal to that of the case where RSDs are included. MG constraints suffer greatly from the loss of RSDs. The LSS probe is no longer able to provide an orthogonal constraint on deviations from GR and the final joint constraint with cross-correlations is over an order of magnitude smaller in terms of MG FoM than when RSDs are included.

In the following sub-sections we perturb in turn a range of the assumptions we have made for our fiducial forecast. Some of the results are summarised in table \ref{tab:summary} for ease of comparison.

\subsection{Survey Specific Target Selection}
\label{sec:steph_nz}
We have assumed a generic survey strategy for our spec-z survey in order that our results might be as widely applicable as possible. Another benefit of assuming a constant galaxy density out to $z=1.7$ is that our results are not hostage to the particularities of a certain target selection strategy which could interact in a peculiar way with, for example, redshift coverage, photo-z error or k/z-dependence of galaxy bias.

We recognise that no individual survey as carried out will look exactly like the toy survey we assume here. The impact of specific survey and target selection strategies are considered in our companion paper \citet{jouvelea_2013}. In this section we reproduce the constraints of one representative survey drawn from that paper to illustrate the differences with respect to the survey model we assume here. We choose a 5000 deg$^2$ survey with an exposure time of 20 minutes using a 4000 fibre spectrograph with a 3 deg$^2$ field of view. We assume two galaxy populations, LRGs and ELGs are separately targeted in a ratio of 30/70. Assuming 8 hours observing per night and a 10\% overhead in survey time, we estimate that it would take about 139 observing nights to saturate the available target list.

\begin{figure}
  \begin{flushleft}
    \centering
       \includegraphics[width=3in,height=3in]{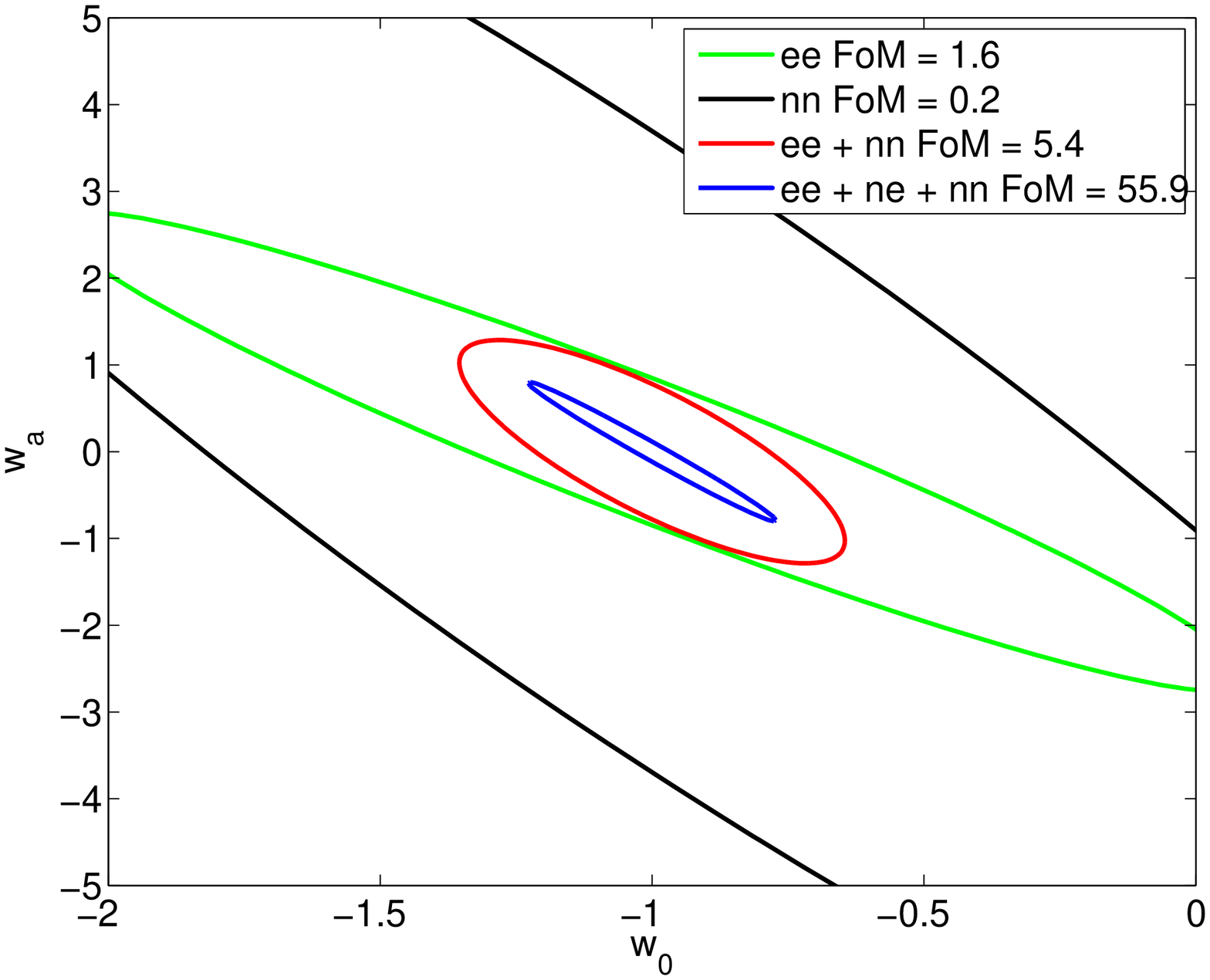}
       \includegraphics[width=3in,height=3in]{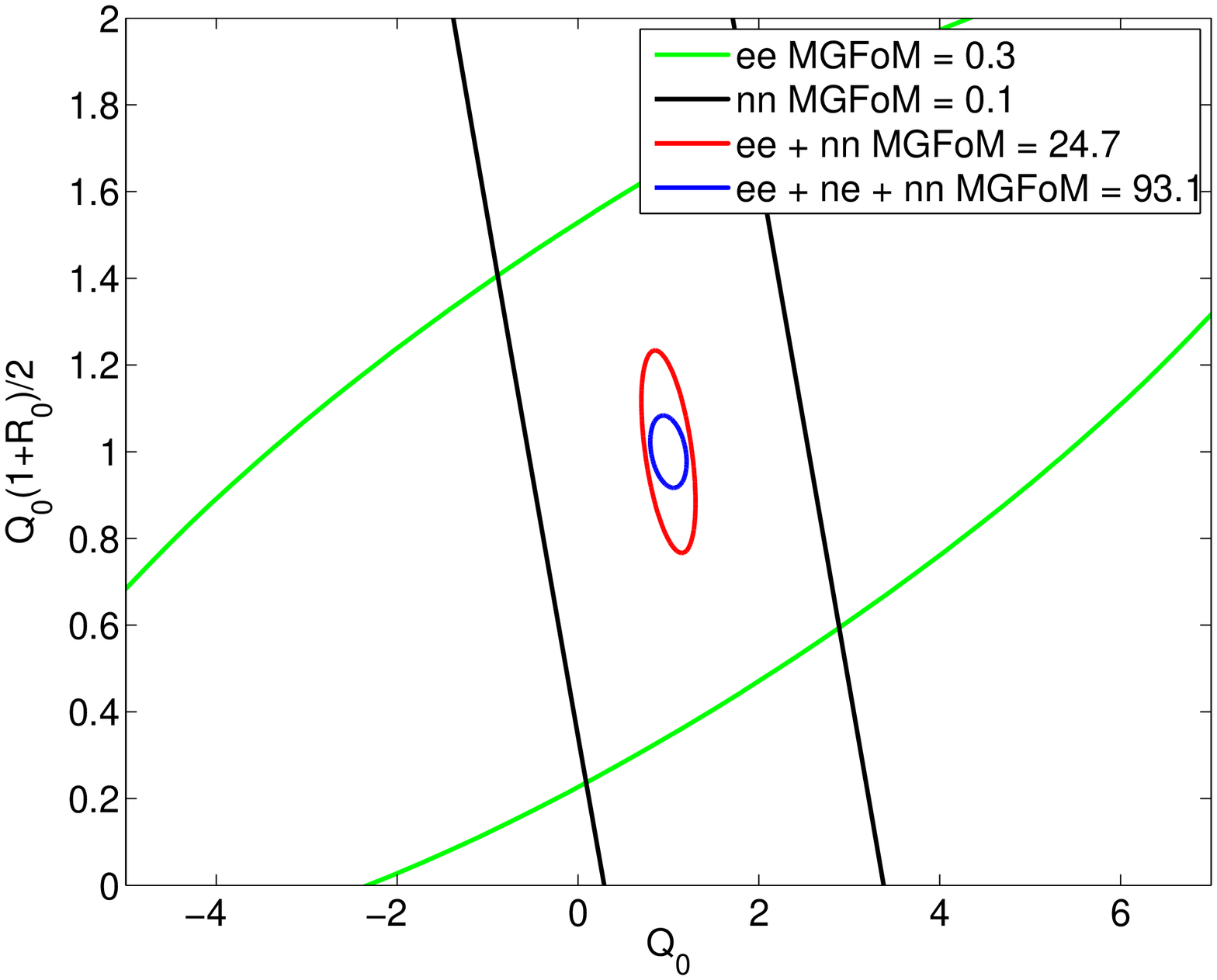}
\vspace{6mm}
\caption{Same forecast assumptions and layout as Fig \ref{fig:contours_fid} but here we assume a specific target selection and survey strategy for the spec-z survey. We assume a $5000$ deg$^2$ survey with 300 observing nights, 20min exposure time and a 30/70 split between LRG/ELG targets. All other assumptions are consistent with our fiducial spec-z survey. The WGL observables come from our fiducial photo-z model survey. For more details see \citet{jouvelea_2013}.}
\label{fig:contours_fid_steph}
  \end{flushleft}
\end{figure}


Despite the differences in survey strategy between our toy model and this more specific example, the basic trends in DE \& MG constraints from spec-z/photo-z combinations are relatively robust. While the $nn$ survey alone is less constraining due to a more uneven z-distribution and slightly smaller z-range, the independent $\epsilon\epsilon + nn$ combination continues to improve on the WGL-alone constraint by more than a factor of three in the case of DE and nearly two orders of magnitude in the case of MG. The same-sky benefit from including the $n\epsilon$ correlations is more pronounced with these specific survey assumptions. In fact the joint DE constraining power is better than for our fiducial scenario while the joint MG result is lower but roughly similar. What is clear is that none of the simplifying assumptions we have made in our fiducial scenario badly bias the trends we are interested in exploring when our spec-z and photo-z surveys are combined. Nevertheless we will continue to use the simple fiducial spec-z survey scenario so that the other assumptions we explore below can be quantified without a complicated interplay with the irregular n(z) that is the result of some specific target selection assumptions.

\begin{table*}
   \centering
   \begin{tabular}{@{} l|c|c|c|c|c|c|c|c|c|c @{}} 
   \hline
Scenario & \multicolumn{2}{|c|}{Photo-z} & \multicolumn{2}{|c|}{Spec-z} & \multicolumn{2}{|c|}{Photo-z + Spec-z} & \multicolumn{2}{|c|}{Photo-z $\times$ Spec-z} & \multicolumn{2}{|c|}{Same-sky Benefit} \\
 & DE & MG & DE & MG & DE & MG & DE & MG & DE & MG \\
\hline
Fiducial 		  & 1.6 & 0.3 & 0.9 & 0.6 & 7.2 & 50.1 & 27.8 & 112.7 & 3.9 & 2.25 \\
`Survey' $n(z)$  & 1.6 & 0.3 & 0.2 & 0.1 & 5.4 & 24.7 & 55.9 & 93.1 & 10.4 & 3.8 \\
With Planck       & 10.2 & 2.1 & 3.9 & 2.2 & 14.0 & 72.4 & 34.7 & 167.0 & 2.5 & 2.3 \\
Marginalise $r_g$ & 1.6 & 0.3 & 0.9 & 0.6 & 7.2 & 50.1 & 8.5 & 56.6 & 1.2 & 1.1 \\
Fix $b_g$ 		  & 1.6 & 0.3 & 6.9 & 4.2 & 16.4 & 86.5 & 48.8 & 185.6 & 3.0 & 2.1 \\
No RSDs          & 1.6 & 0.3 & 0.0 & 0.0 & 5.7 & 1.3 & 26.9 & 7.0 & 4.7 & 5.4 \\
Fix $\delta_z$ 	  & 3.5 & 0.4 & 0.9 & 0.6 & 10.5 & 70.0 & 31.8 & 134.4 & 3.0 & 1.9 \\
$b_g = 2$ 		  & 1.6 & 0.3 & 1.6 & 1.1 & 9.6 & 66.2 & 73.9 & 159.0 & 7.7 & 2.4 \\
15,000 deg$^2$ spec-z  & 1.6 & 0.3 & 2.7 & 12.6 & 11.3 & 201.9 & 38.6 & 423.4 & 3.4 & 2.1 \\
\hline
   \end{tabular}
   \caption{Summary of constraints from our photo-z and spec-z surveys for a number of forecast assumptions. We show DE \& MG FoMs for Photo-z $\epsilon\epsilon$, Spec-z $nn$, Photo-z + Spec-z $\epsilon\epsilon+nn$ and Photo-z $\times$ Spec-z $\epsilon\epsilon+n\epsilon+nn$, where the $\epsilon$ observables are always drawn from the photo-z survey and $n$ are always drawn from the spec-z survey. We also show the Same-sky Benefit for DE \& MG i.e. the FoM for $\epsilon\epsilon+n\epsilon+nn$, divided by that for $\epsilon\epsilon+nn$. We show the results for our fiducial surveys and fiducial assumptions and then perturb one assumption at a time: using a particular target selection/survey strategy scenario from \citet{jouvelea_2013}; including Planck forecast priors; marginalising over $r_g$ the same way we do over $b_g$; Fixing both $b_g$ and $r_g$; removing RSDs from our $n$ observables; fixing the photometric redshift uncertainty, $\delta_z$; changing the fiducial galaxy bias amplitude to $b_{g}=2$, increasing the area of the spec-z survey to 15,000 deg$^2$ (with the same no.density per deg$^2$) while keeping the area of the photo-z survey fixed at 5,000 deg$^2$. Each change is made independently, all other assumptions are fixed at those of our fiducial forecasts. Details of these scenarios can be found throughout section \ref{sec:forecasts}.}
   \label{tab:summary}
\end{table*}

\subsection{CMB Prior}

This paper is primarily concerned with the details of the combination of a spec-z galaxy clustering/RSD survey with a photo-z WGL survey. It is from this sort of joint probes analysis that all the most stringent constraints on cosmology will be derived. Of course there are cosmological probes beyond LSS \& WGL and any comprehensive constraints on cosmology will have to integrate them into the analysis.

When adding extra cosmological observables to our analysis the most useful are those that provide orthogonal constraints on cosmological parameters, thereby breaking degeneracies present in the analysis and improving the final results. Adding ``disjoint'' cosmological probes which are sensitive to different physics is a powerful way of breaking degeneracies in particular observables. In our case LSS and WGL are both late-Universe probes sensitive to the growth of structure as the Universe evolves. Two of the most useful probes to add to this mix are therefore type Ia supernovae (SNe) and the Cosmic Microwave Background (CMB). The CMB probes the physics of the early Universe while SNe directly constrain the expansion history of the Universe (it should be noted that WGL also has some direct access to expansion history through its geometric kernel). 
The best CMB observations to date come from the Planck satellite \citep{planck1}. We have reproduced our fiducial results from Fig. \ref{fig:contours_fid} including a prior based on a forecast for the Planck mission [Dark Energy Survey Theory \& Combined Probes group, private communication]. It would have been more complete to include the newly released Planck results themselves in our combined data but would have delayed the release of this paper. We are informed [Tom Kitching, private communication] that the released Planck constraints and forecasts are similar enough for our present needs. 

Clearly the addition of the Planck constraints means that each of our previous probe combinations is correspondingly more powerful than they were on their own, see Table \ref{tab:summary} for details. The strongest improvements come for the individual probes with, for the case of DE, $\epsilon\epsilon$ improving by a factor of 6 and $nn$ by a factor of more than 4. What is striking is how the combination of the WGL \& LSS surveys, with and without cross-correlation, still offer significant improvements through the breaking of parameter degeneracies. While the independent, $\epsilon\epsilon + nn + CMB$, combination is only $\tilde 40\%$ more powerful than $\epsilon\epsilon + CMB$, the addition of the $n\epsilon$ cross-correlation produces a same-sky benefit factor of $\sim$2.5. This is lower than the case without Planck but still a substantial benefit. The MG same-sky benefit is robust to the inclusion of Planck, unsurprising as the CMB can tell us little about late-time modifications to GR.





\subsection{The Importance of Galaxy Bias}
\label{sec:sec2}

The primary nuisance parameter we are interested in is galaxy bias, $b_{g}(k,z)$, which accounts for the fact that galaxies are a biased tracer of the underlying dark matter distribution. We have assumed a linear galaxy bias model,

\begin{equation}
\delta_g = b(k,z) \delta_m,
\end{equation}
where the galaxy overdensity, $\delta_g$, is related to the matter overdensity, $\delta_m$, by a single function of scale and redshift. This propagates to a simple relation between the galaxy power spectrum and the underlying DM power spectrum,
\begin{equation}
P_{gg}(k,z) = b^{2}(k,z) P_{mm}(k,z).
\end{equation}

Our default parameterisation of this function allows a single overall amplitude and variation in k/z-space through modulation of $2 \times 2$ grid nodes covering our full redshift range and all linear/quasi-linear scales. Galaxy bias is not particularly well understood or empirically constrained at present, especially on non-linear or quasi-linear scales \citep{marinea_wigglez3pt_2013,contreras_wiggle2pt_2013,comparat_bias_2013,pujol_gaztanaga_2013}. It is this which motivates our stringent cuts in ell (see section \ref{sec:kcuts} for more details). Even so, the assumptions we make on galaxy bias, even at linear scales, can dramatically effect the constraining power of our spec-z survey and the usefulness of same-sky correlations. These effects will be investigate din more detail in Clerkin et al. (in prep).  

In this section we explore the impact of these assumptions on our constraints from the spec-z LSS survey and its combination with our photo-z WGL survey. We divide our assumptions about galaxy bias into three: the underlying model of $b_g$ which we assume; the fiducial amplitude of $b_g$ for the galaxy population captured by our survey and the level of uncertainty we allow to enter our $b_g$ model in the form of nuisance parameters. We finish the section by exploring the impact of a related quantity, $r_g$, the galaxy-shear cross-correlation coefficient.  

\subsubsection{Galaxy Bias Model}
\label{sec:bg_models}

For our fiducial model we assume that galaxy bias varies around unity in some redshift- and scale-dependent way parameterised by our $1 + n_k \times n_z$ nuisance parameters (grid nodes in k \& z plus an overall amplitude term). In effect we assume that galaxy bias is constant with scale and redshift but we allow enough uncertainty to encompass the true $b_g$ evolution. Here we consider whether the choice of a fiducial model can change our forecasts.

As well as our fiducial $b_g = 1$ model we also run forecasts assuming galaxy bias is given by the Fry \citep{fry_biasmodel_1996} 
and Q \citep{cole_2005_2df} models. These are two regularly used galaxy bias models, motivated by simple physical arguments and N-body simulations. The Fry model gives a z-dependent $b_g$, while the Q-model produces a k-dependent $b_g$.

The Fry biasing model assumes the continuity equation and and linear growth,
\begin{equation}
b_{g,Fry}(z) = 1 + \frac{b_{0}-1}{D(z)},
\end{equation}
where $D(z)$ is the linear growth function and $b_0$ is known as the ``Fry parameter'', for which we assume a fiducial value $b_0 = 2$.

The Q model is motivated by a need to allow an unknown scale dependence to enter the galaxy bias formalism. The Q-model, derived for low-redshift, once calibrated against N-body simulations populated by a semi-analytic galaxy formation model, takes the form
\begin{equation}
b_{g,Q}(k) = \sqrt{\frac{1+Qk^2}{1+1.4k}},
\end{equation}
where we set $Q=4$ following the usage in \citet{swansonea_2010}. 

We compare results from these different galaxy bias models in the unrealistic case where we assume that our model exactly captures the physics of galaxy bias with zero uncertainty. In this case, chosen for maximum difference between models, there is significant scatter between our results, with the Fry model in particular producing marginalised errors on $w_0, w_a$ which are less than half those of the fiducial model for some parameters. This can be understood because of the extra cosmology dependence encoded in the Fry model, allowing an extra handle on the cosmology we're trying to measure. 

The Q-model produces results in relatively good agreement with the fiducial model but this is an artifact of the stringent ell-cuts we apply to our LSS probes. As the signature features of the k-dependent (z-independent) Q-model only kick in for scales smaller than those we include, their effect is excluded from our forecast. if we cease to apply cuts at quasi-linear scales, trusting our knowledge of non-linear physics up to $l=3000$ (wildly optimistic) then we see large divergence between forecasts assuming our fiducial $b_g$ and those of the Q-model. In this regime the Q-model over-constrains cosmology due to the non-physical nature of its predictions for very small scales. 

The divergence between forecasts for different fiducial galaxy bias models reduces as the number of nuisance parameters increases.

Galaxy bias model choice is an important and involved topic which we only have the space to scratch the surface of in this section. We have demonstrated the need to pay attention to $b_g$ modelling when quoting results of LSS or combined surveys. We are content that a sufficient number of nuisance parameters can dilute the difference between models and that our fiducial nuisance parameterisation is sufficient in this respect. 

\subsubsection{Galaxy Bias Amplitude}
\label{sec:bg_amplitude}

Recently there has been some attention in the literature to the use of multiple, differently biased galaxy populations which can be used simultaneously as tracers of LSS in such a way as to reduce cosmic variance \citep{mcdonald_seljak_2009,bernstein_cai_2011}. 
A full implementation of this approach would consider a number of differently biased galaxy populations and calculate $C(l)$s for their auto- and cross-correlations in all the available tomographic bin pairs. Each galaxy population would require its own set of nuisance parameters which could be ``self-calibrated'' through cross-correlation with the photo-z survey.

While such a full implementation of this approach is beyond the scope of this paper we want to underline the importance of bias amplitude through a simple example.

We perform our fiducial forecast again but assuming that we have targeted a different, more strongly biased population of galaxies, setting  $b_g = 2$ and assuming the standard five nuisance parameters to allow uncertainty around this new amplitude.

As expected the spec-z survey alone is more constraining, by a factor of 1.8 in DE FoM. Interestingly while the independent non-overalpping combination of photo-z and spec-z surveys is slightly improved (a factor of 1.3), the same-sky combination sees strong improvement by a factor of 2.6 in DE FoM compared to the $b_{g}=2$ case. This means that the same-sky benefit factor goes from 3.9 with $b_g=1$ to 7.7 with $b_g=2$ all thanks to the greater signal-to-noise that comes from targeting more highly biased tracers.

The trends in the MG case are similar but markedly less pronounced, with a change in same-sky benefit factor from 2.25 with $b_g=1$ tot 2.4 with $b_g = 2$.

\subsubsection{Galaxy Bias Nuisance Parameterisation}
\label{sec:bg_nknz}

\begin{figure*}
  \begin{flushleft}
    \centering
       \includegraphics[width=6in,height=4in]{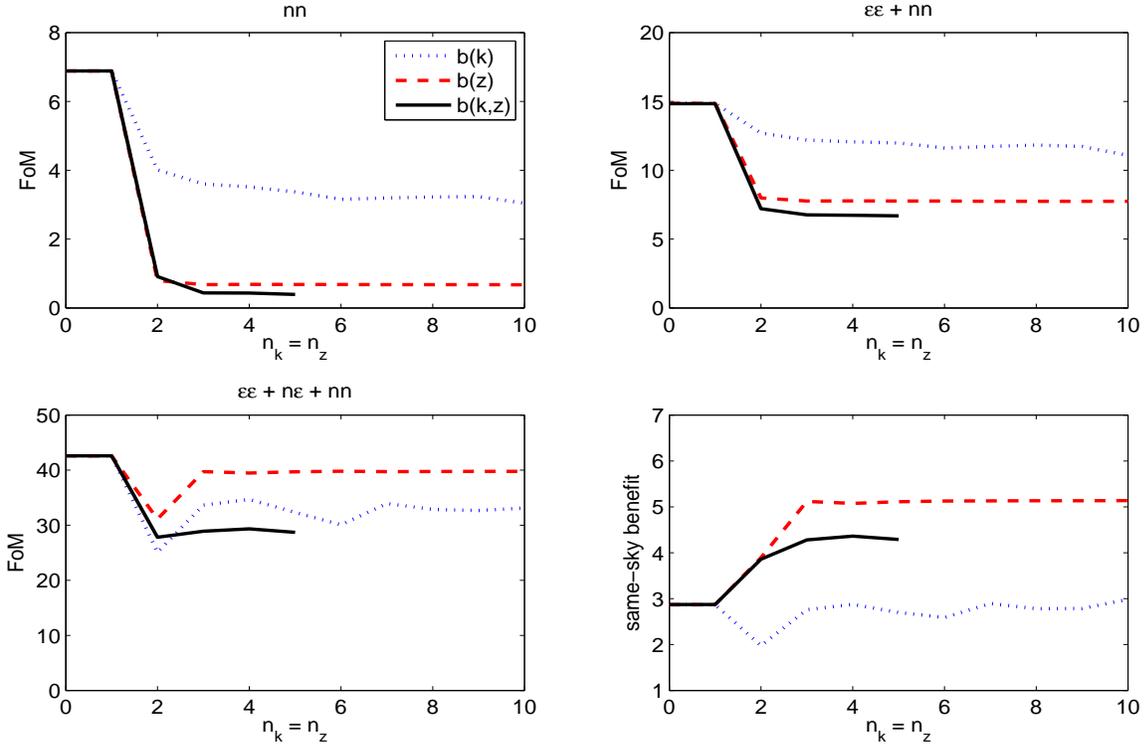}
\vspace{6mm}
\caption{ DE FoM as a function of the number of grid nodes in k/z space for our galaxy bias model. Fiducial surveys and forecast assumptions assumed. Results are shown for $nn$-only [top left], $\epsilon\epsilon + nn$ [top right] and $\epsilon\epsilon + n\epsilon + nn$ [bottom left]. We also show the relative improvement due to a same sky analysis [bottom right] i.e. the ratio of $FoM_{\epsilon\epsilon + n\epsilon + nn}$ to $FoM_{\epsilon\epsilon + nn}$. Each panel presents results for bias which depends only on redshift, $b_{g}(z)$, [red dashed], bias which depends only on scale, $b_{g}(k)$, [blue dotted], and bias which depends on both scale and redshift, $b_{g}(k,z)$, [black solid]. $n_k$ and $n_z$ are the number of nodes in k/z space. These vary around their fiducial value of one and are interpolated to give a $b_g$ ``surface'' in k/z. z-nodes are linearly spaced, k-nodes are log-spaced.}
\label{fig:bg_kz}
  \end{flushleft}
\end{figure*}

Fig. \ref{fig:bg_kz} shows DE FoM as a function of the number of grid nodes marginalised over in $b_{g}(k,z)$, showing spearate results for scale-dependent, $b_{g}(k)$, redshift-dependent, $b_{g}(z)$, and both scale- and redshift-dependent galaxy bias, $b_{g}(k,z)$. More grid nodes means more flexibility and reduces the constraining power of our spec-z survey. The $b_{g}(k,z)$ results stop at $n_k = n_z = 5$ for computational reasons. Our implementation is based on that of \citet{joachimi_bridle_2009}.

The top-left panel shows results for the spec-z $nn$ survey alone. FoM decreases as we increase the flexibility of the grid. While FoM falls quickly to very low levels for $b_{g}(k,z)$ \& $b_{g}(z)$, the $b_{g}(k)$ FoM retains roughly half it's max value by $n_k = 2$ and falls slowly thereafter, reaching 1/3 of its maximum value by $n_k = 10$. It's possible that this relative insensitivity to an increaingly scale-dependent redshift term is due to the stringent ell-cuts we impose on our spec-z survey, removing non-linear and quasi-linear scales. 

Interestingly, there is a relatively consistent plateau above $n_k = n_z = 2$ for all bias types and all probe combinations, suggesting our fiducial $2 \times 2$ grid approach is a sensible choice if we are not to over-constrain cosmology from LSS. A higher resolution grid is computationally more intensive without significantly affecting the cosmological constraints produced. In particular we tested the z-dependent bias, $b_{g}(z)$, out to 40 nuisance parameters (one for each tomographic bin) and found negligible decrease in constraining power as compared to the $n_z = 10$ case.

The photo-z WGL survey is insensitive to galaxy bias. When it is included, either with or without cross-correlations, we not only see an overall improvement in FoM but there is a ``floor'' below which the FoM does not fall with increased grid flexibility, this can be thought of as the residual constraining power of the combination after $nn$ has been marginalised out of existence. It is interesting that this floor is lower for the $b_{g}(k,z)$ case than for the $b_{g}(k)$ case, wtith $b_{g}(z)$ lying only slightly above the $b_{g}(k,z)$ line. We expect $b_{g}(k)$ to perform best in combination because it retains most information for $nn$-alone. 

The bottom left panel shows results in the case of the full same-sky combination $\epsilon\epsilon + n\epsilon + nn$. That each of the bias cases shows relative stability in the face of increased model flexibility demonstrates the power of the cross-correlations to control unknown bias terms. The trend with increased number of grid nodes is less smooth than in the other cases. This is not particularly unexpected- as the number of grid nodes is changed, so is their spacing in k/z so we would not require the FoMs produced to change monotonically. Nevertheless this result does suggest that the exact location of our $b(k,z)$ ``flexibility'' can influence our results and should be treated with care. Our fiducial choice is conservative enough that we are not over-estimating our understanding of galaxy bias, we are into the regime where the survey is `calibrating itself', increasing the number of nuisance parameters would not affect our results overly.

The same-sky benefit results (bottom right panel) correspond to the trend in the other plots. Same-sky benefit improves with increased uncertainty in $b_g$ as the cross-correlations act to calibrate the galaxy bias. The trend is least pronounced in the case of $b_{g}(k)$ where the impact of increased bias uncertainty is limited. It should be noted that even in the case of a fixed $b_g$ (i.e. the un-justified assumption that we understand the bias term perfectly) there is still significant, $\sim \times 3$,
improvement from same-sky, confirming that the cross-correlation's effects are not limited to better control of galaxy bias uncertainty.




\subsubsection{$r_g$, the cross-correlation coefficient}
\label{sec:bg_rg}
Just as galaxy bias, $b_{g}(k,z)$, is a nuisance parameter which describes our ignorance of the extent to which galaxies are a biased tracer of dark matter, there is an analogous term in the $n\epsilon$ observable which appears where we cross-correlate galaxy clustering and cosmic shear. The term, which we refer to here as the galaxy-shear cross-correlation coefficient, $r_{g}(k,z)$, is a measure of the statistical coherence of the two fields \citep{baldauf_rg_2010,guzik_seljak_2001_rg, mandelbaumea_2013}. $r_{g}=1$ means the fields (galaxy and matter overdensities respectively) are fully correlated, there is a deterministic mapping between the two fields.
\citet{gaztanaga_pau_2012} have plausibly argued that, when we restrict ourselves to linear scales, where galaxy bias can be assumed to be broadly scale-independent, then we can assume $r_{g}(k,z) = 1$. \citet{cacciato_lahav_bosch_hoekstra_dekel_2012} found $r_{g} \approx 1$ on large scales based on the halo model.

In this section we relax this assumption, parameterising $r_{g}(k,z)$ in the same way as we treat $b_{g}(k,z)$, i.e. one free amplitude term and a $2 \times 2$ grid in k/z-space, and marginalising the resulting five nuisance parameters.

Only the cross-correlation term, $n\epsilon$, is sensitive to $r_{g}(k,z)$. As such, marginalisation over this extra term only effects our same-sky combination of probes, $\epsilon\epsilon + n\epsilon + nn$, reducing the benefit from a same sky analsysis. Assuming, as we do here, that $r_g$ is as strong a contaminant as $b_g$ is a rather pessimistic scenario, strongly penalising same-sky coverage. Even so there is still a $\sim 25\%$ improvement when same-sky constraints are compared to the independent combination of surveys. 

\subsection{Photometric Redshift Error}
\label{sec:photoz_error}

As well as improvements in the ability to constrain cosmology and control for galaxy bias, there has been much interest in the combination of photo-z and spec-z surveys to ``self-calibrate'' the photometric redshift error \citep{newman_2008_photoz,zhang09}. The principle being invoked here is straightforward: if the spec-z survey offers us highly accurate redshifts for some sub-set of the galaxies in the photo-z survey then we should be able to use this information to learn more about the photometric redshift distribution than we can using the photo-z survey alone. 

In practice the extra specroscopic redshift information could be fundamentally integrated into the calibration of the photo-z sample. Here we take a more general approach parameterise the error on our photometric $n(z)$ in some way, then allow these parameters to vary as new nuisance parameters. Extra photo-z `'calibration'' from the addition of the spec-z survey enters as tighter constraints on these photo-z nuisance parameters. This information is only available when the two surveys overlap so it forms a contribution towards the same-sky benefit we observe.

In our fiducial model we use a single global parameter $\delta_z$ as our photo-z nuisance parameter. It enters into the overall photo-z error as $\sigma_z = \delta_z (1+z)$ and we allow it to vary around our DES-like fiducial value of $\delta_z = 0.07$. One photo-z nuisance parameter is relatively conservative but as there have been some reservations expressed about the efficacy of this kind of ``self-calibration'' we consider this a conservative choice. Having many photo-z nuisance parameters weights the entire forecast methodology strongly towards a poorly estimated photo-z distribution which becomes much improved by cross-correlation with a spec-z survey. 

In this section we relax some of these assumptions and investigate in more detail the impact of photo-z mis-estimation on our survey constraints and the same-sky benefit from cross-correlation.

We follow the approach of \citet{bordoloi_2010,amarar07} by introducing $2N_z$ nuisance parameters, where $N_z$ is the number of tomographic bins used to analyse our photo-z survey ($N_z = 5$ for our DES-like survey). We allow $\delta_z$ to vary independently in each z-bin around the fiducial value of 0.07 and we introduce a new nuisance parameter- the bias on the mean redshift of each bin, which we allow to vary independently around zero. We allow these ten nuisance parameters to vary freely with wide, flat priors. This represents a case of very poor photo-z estimation. We then increase the prior on all the photo-z nuisance parameters to show the change in FoMs and same-sky benefit with improving photo-z knowledge. Results are shown in Fig. \ref{fig:partial_overlap}. 

\begin{figure}
  \begin{flushleft}
    \centering
       \includegraphics[width=3in,height=2in]{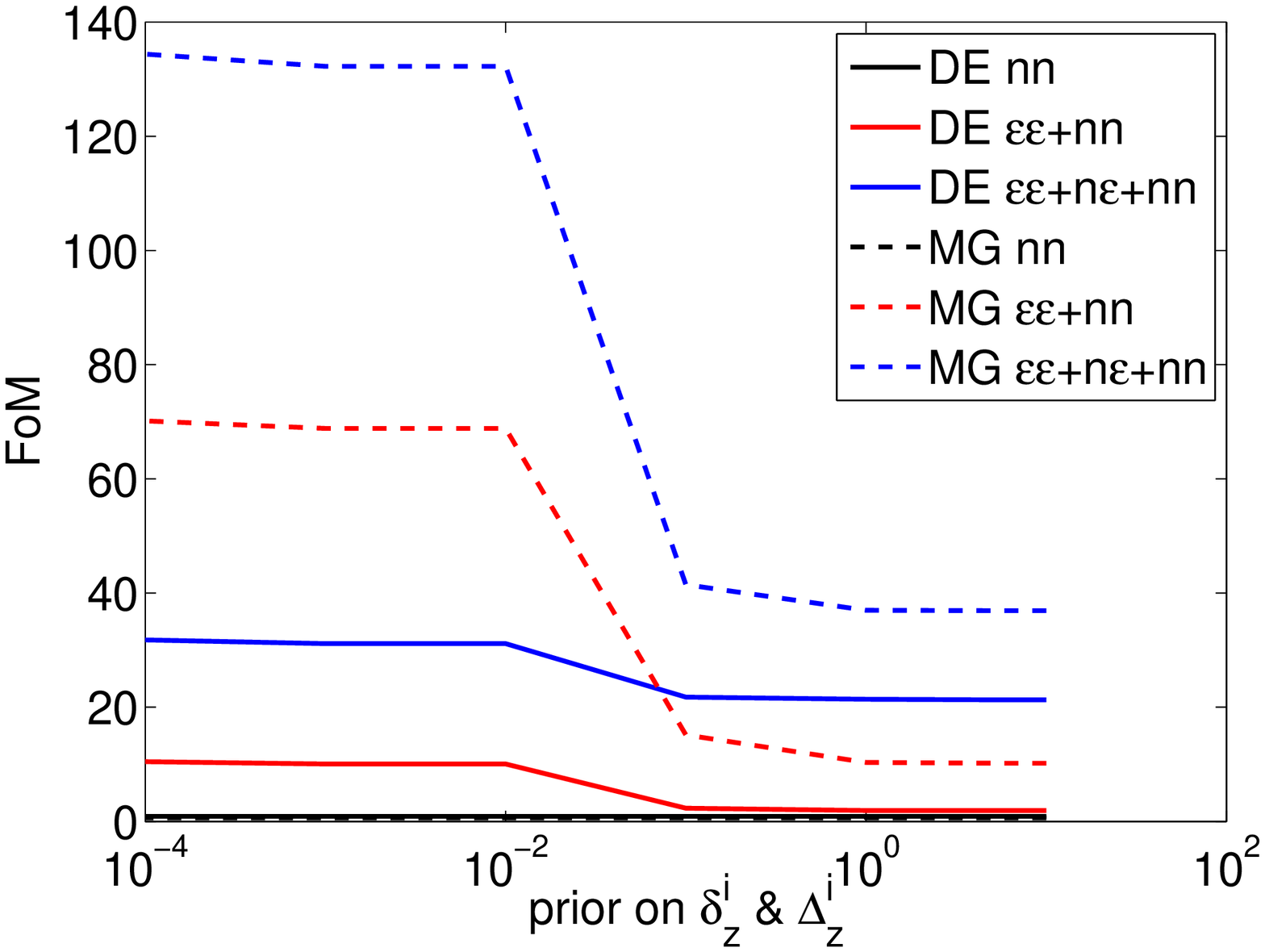}
       \includegraphics[width=3in,height=2in]{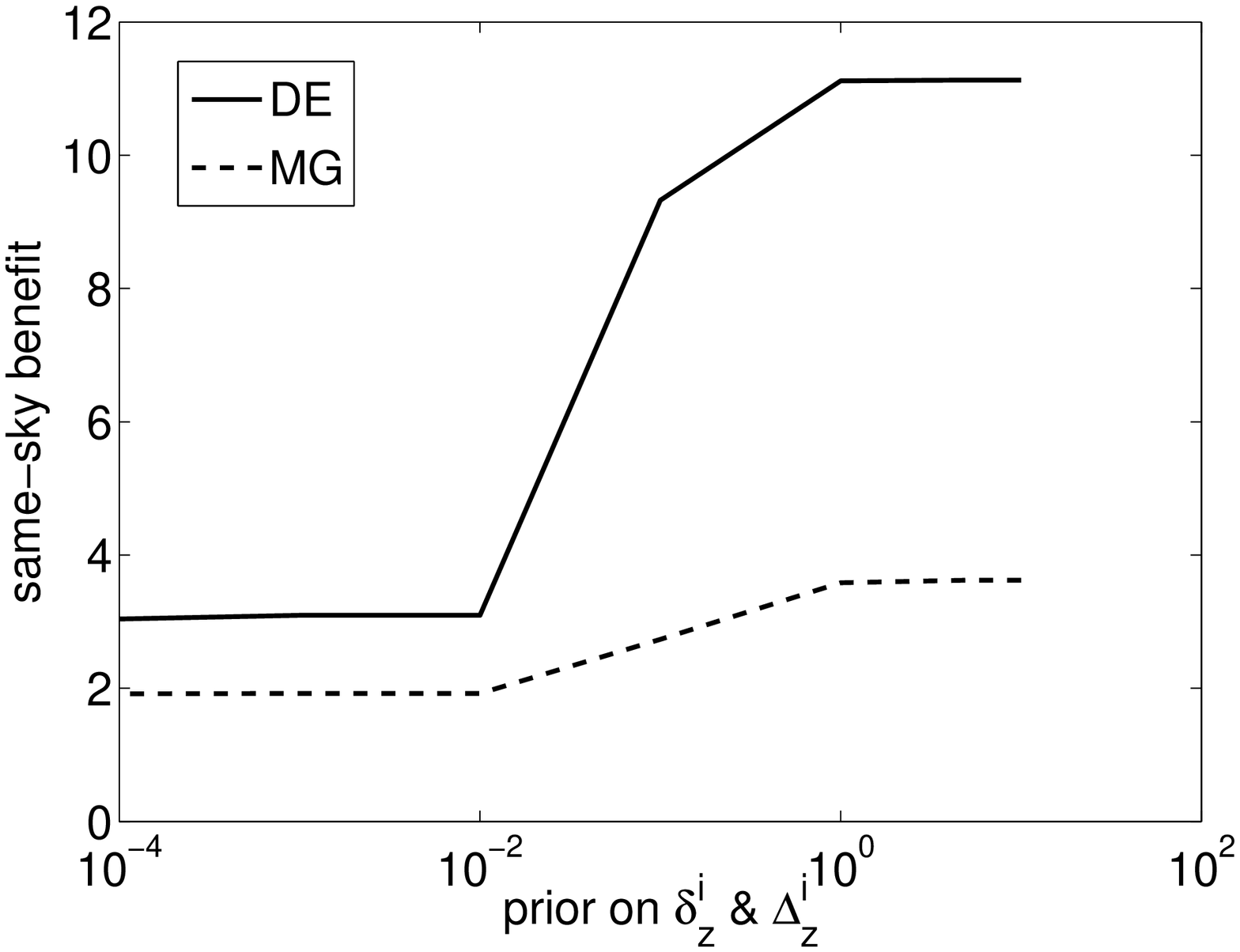}
\vspace{6mm}
\caption{This plot uses our fiducial forecast assumptions and surveys but allows photo-z error to vary more freely. We use ten photo-z nuisance parameters: a gaussian error per tomographic bin, $\delta_{z}^{i}$, with fiducial values 0.07, and a mean redshift offset per bin, $\Delta_{z}^{i}$, with fiducial values 0. [Top panel] DE [solid lines] \& MG [dashed lines] FoMs as a function of the prior on our photometric redshift nuisance parameters, $\delta_{z}^{i}$ \& $\Delta_{z}^{i}$. FoMs are shown for $nn$ alone [black lines], $\epsilon\epsilon+nn$ [red lines] and $\epsilon\epsilon+n\epsilon+nn$ [blue lines]. [Lower panel] Same-sky benefit as a function of prior on $\delta_{z}^{i}$ \& $\Delta_{z}^{i}$ for DE [solid] and MG [dashed].}
\label{fig:partial_overlap}
  \end{flushleft}
\end{figure}

In the case of wide, flat prior (right hand side of the plots) FoM for both DE and MG is reduced for any probe combination that includes the $\epsilon\epsilon$ photo-z survey (the $nn$ lines are flat as the photo-z nuisance parameters do not impact this probe). Moving from right to left on the plot, the priors on all the photo-z nuisance parameters are tightened. This improves the constraining power of any survey including $\epsilon\epsilon$ as marginalising over the photo-z nuisance parameters has less impact. The improvement is more pronounced for MG than it is for DE, suggesting that the z-dependence of the MG parameters is more degenerate with the photo-z error than that of $w_0,w_a$. The major improvement in constraining power comes between prior values of 0.1 and 0.01, with plateaus above and below this range. This finding concurs with that of \citet{MGPaper2}.

As prior ranges on the photo-z nuisance parameters are tighterned, the same-sky benefit drops (lower panel). This is expected as one effect of the $n\epsilon$ cross-correlation is to ``self-calibrate'' photo-z error. As the priors are tightened, the cross-correlation has less work to do so the benefit it confers is less. Even so, while the DE same-sky benefit reduces from a high of a factor of 11 for a prior of width 10, it is still more than a factor of 3 with a very tight prior of width $1e^{-4}$. This confirms that the same-sky benefit is not due primarily, or even predominantly, to the self-calibration of photo-z error. In fact, for both DE and MG, the same-sky benefit is relatively stable below a prior of 0.01. 





\subsection{Survey Overlap: Area on the Sky}
\label{sec:partial_overlap}

\begin{figure}
  \begin{flushleft}
    \centering
       \includegraphics[width=3in,height=2in]{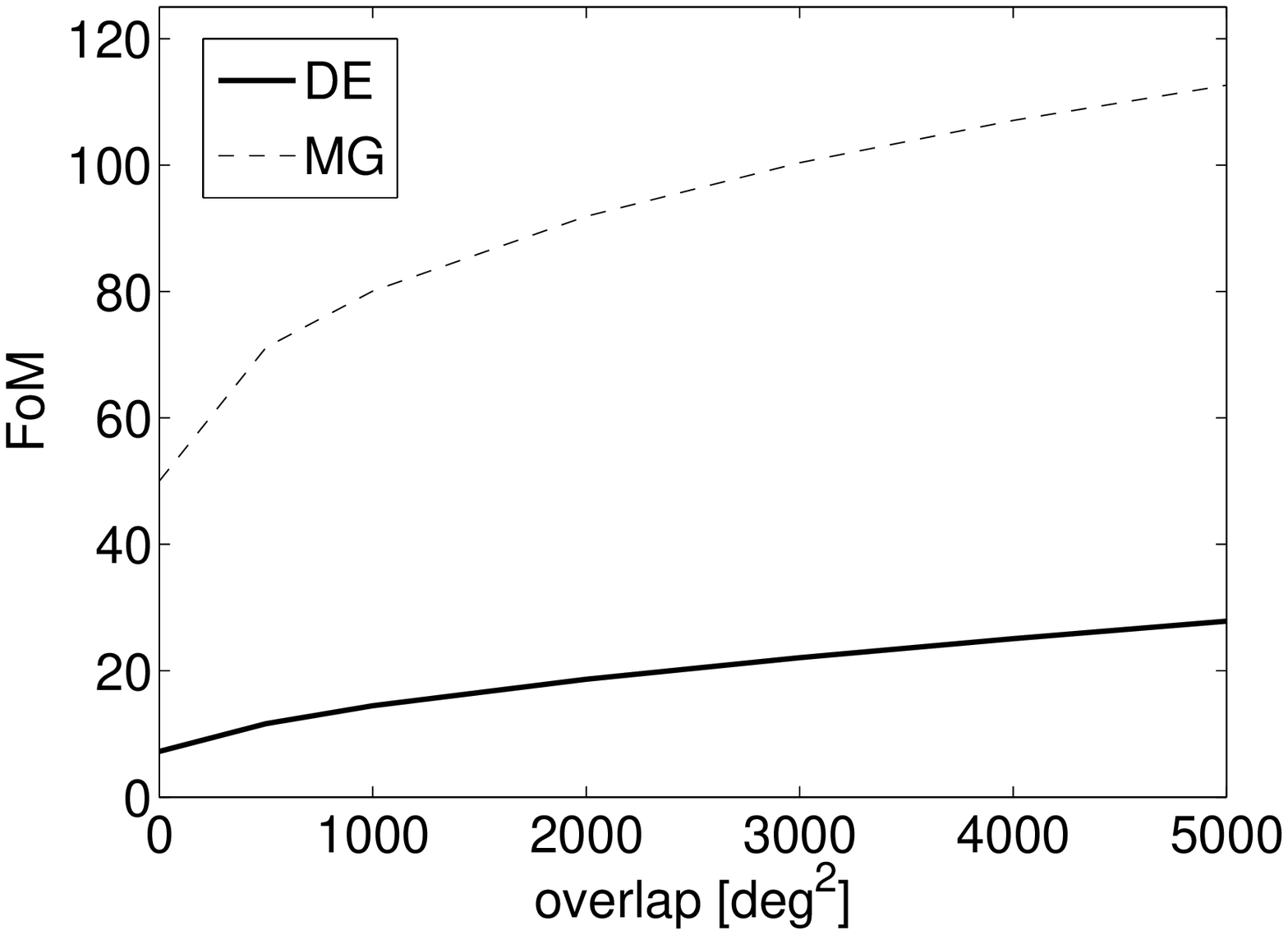}
       \includegraphics[width=3in,height=2in]{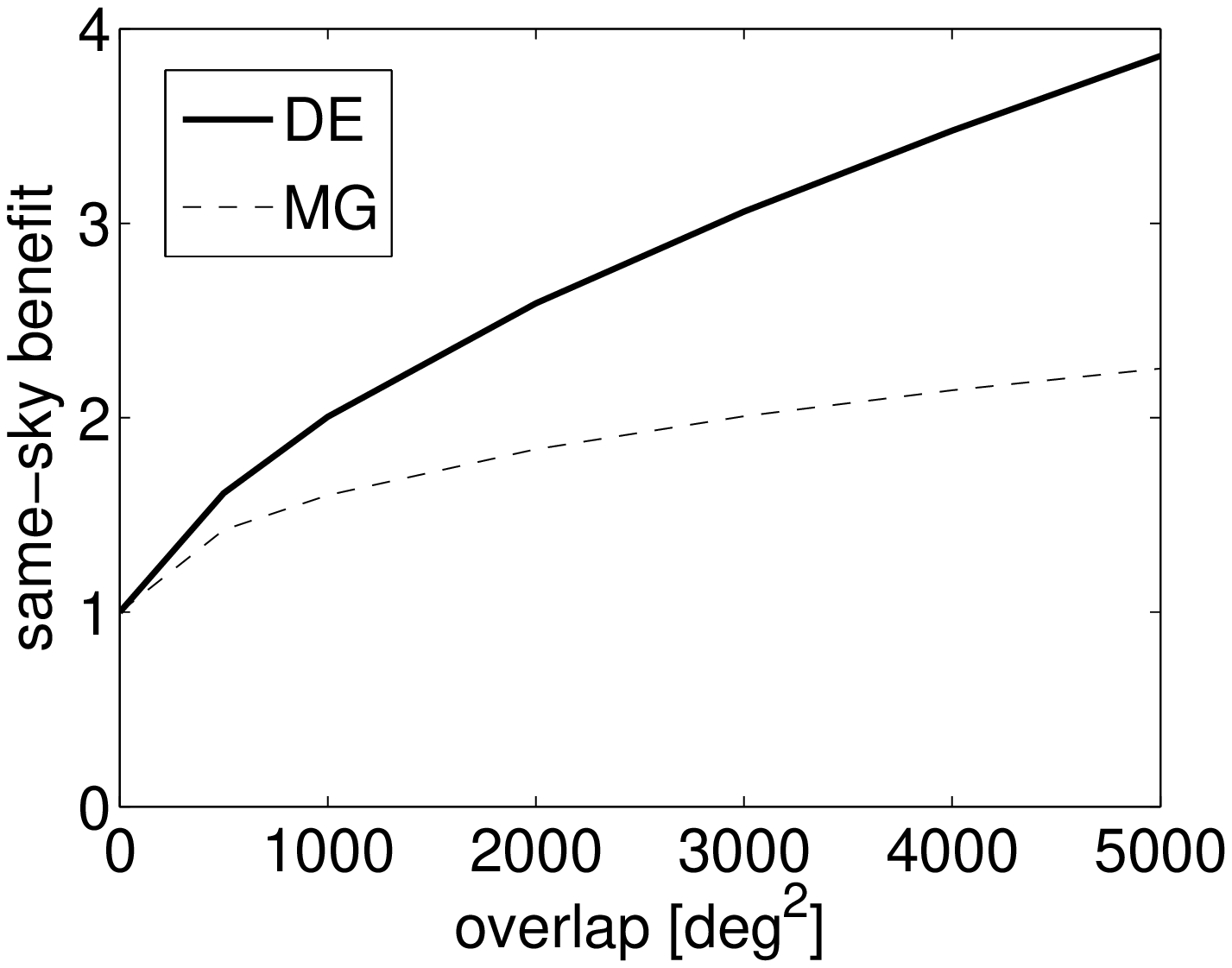}
\vspace{6mm}
\caption{[Top panel] FoMs for DE [solid] and MG [dashed] for our $\epsilon\epsilon + n\epsilon + nn$ from our photo-z and spec-z surveys with fiducial forecast assumptions as a function of survey overlap. Both surveys have their fiducial 5000deg$^2$ area but the $n\epsilon$ cross-correlation can only exploit sky area for which the surveys overlap. For 0 deg$^2$ overlap there is no $n\epsilon$ contribution. [Lower panel] Same-sky benefit i.e. $FoM_{\epsilon\epsilon+n\epsilon+nn}/FoM_{\epsilon\epsilon+nn}$ as a function of overlapping area.}
\label{fig:partial_overlap}
  \end{flushleft}
\end{figure}


The same-sky benefit of overlapping a spec-z LSS and photo-z WGL survey had been a main focus of investigation for this paper. Here and in the next section we examine the results of survey overlap in more detail. In this section we show, in Fig. \ref{fig:partial_overlap}, the relative change in DE FoM as we increase the overlap fraction of our fiducial surveys from zero to a full overlap where the 5000 deg$^2$ of each survey are totally coincident. 

We have already established that the extra observable $n\epsilon$, which is only accessible on patches of the sky where the surveys overlap, improves our ability to constrain cosmology due to its different cosmological and redshift dependence. It is not surprising that increased overlap area improves the FoM for both DE and MG. It is interesting to note that the trend in DE FoM with overlap area is mildly nonlinear suggesting that even a small amount of overlap should be prioritised. This trend is even more pronounced in the MG case where the first 1000 deg$^2$ of overlap provides more than a third of the improvement in FoM gained from the full 5000 deg$^2$ overlap.

Similar trends are apparent in the plot of same-sky benefit vs. overlap area (normalised for plotting purposes in Fig. \ref{fig:partial_overlap} to 1 for zero overlap). Here it is clear how even a small overlapping area can strongly benefit constraints on deviations from GR. The DE FoM benefits more from overlap area, consistent with our initial results for same-sky benefit in fig \ref{fig:contours_fid} above. Even a 1000 deg$^2$ overlap can double the DE FoM compared to the same surveys on separate patches of the sky.

One of the results we consider in table \ref{tab:summary} is a spec-z survey of 15,000 deg$^2$ combined with our same 5,000 deg$^2$ photo-z survey. We keep the number density of the spec-z survey fixed i.e. we capture three times as many galaxies as our fiducial 5,000 deg$^2$ survey. Of course, the spec-z survey is now three times as constraining in DE, bettering the 5,000 deg$^2$ WGL photo-z survey. The MG constraint increases substantially, becoming a factor of 20 better than the 5,000 deg$^2$ example. What is interesting is that we still see substantial improvement from the addition of the 5,000 deg$^2$ photo-z survey both non-overlapping (x4 DE, x16 MG compared to the 15,000 deg$^2$ spec-z survey alone) and overlapping (x3.4 DE, x2.1 MG compared to non-overlapping). This shows that the degeneracy breaking power of the combined constraints is still important, even with such a powerful spec-z survey.

\subsection{Survey Overlap: Redshift Coverage}
\label{sec:z_overlap}

\begin{figure*}
  \begin{flushleft}
    \centering
       \includegraphics[width=3in,height=2in]{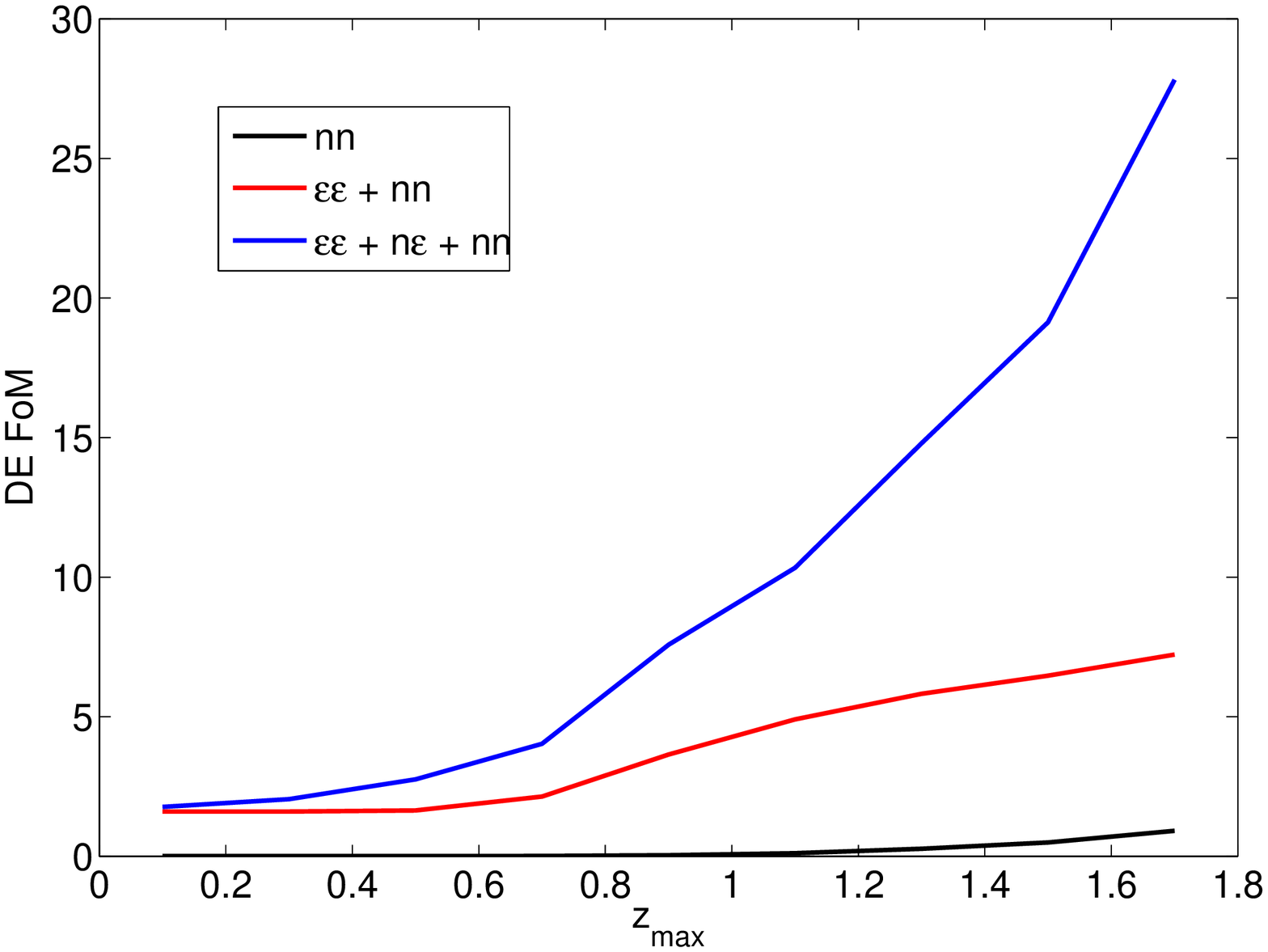}
       \includegraphics[width=3in,height=2in]{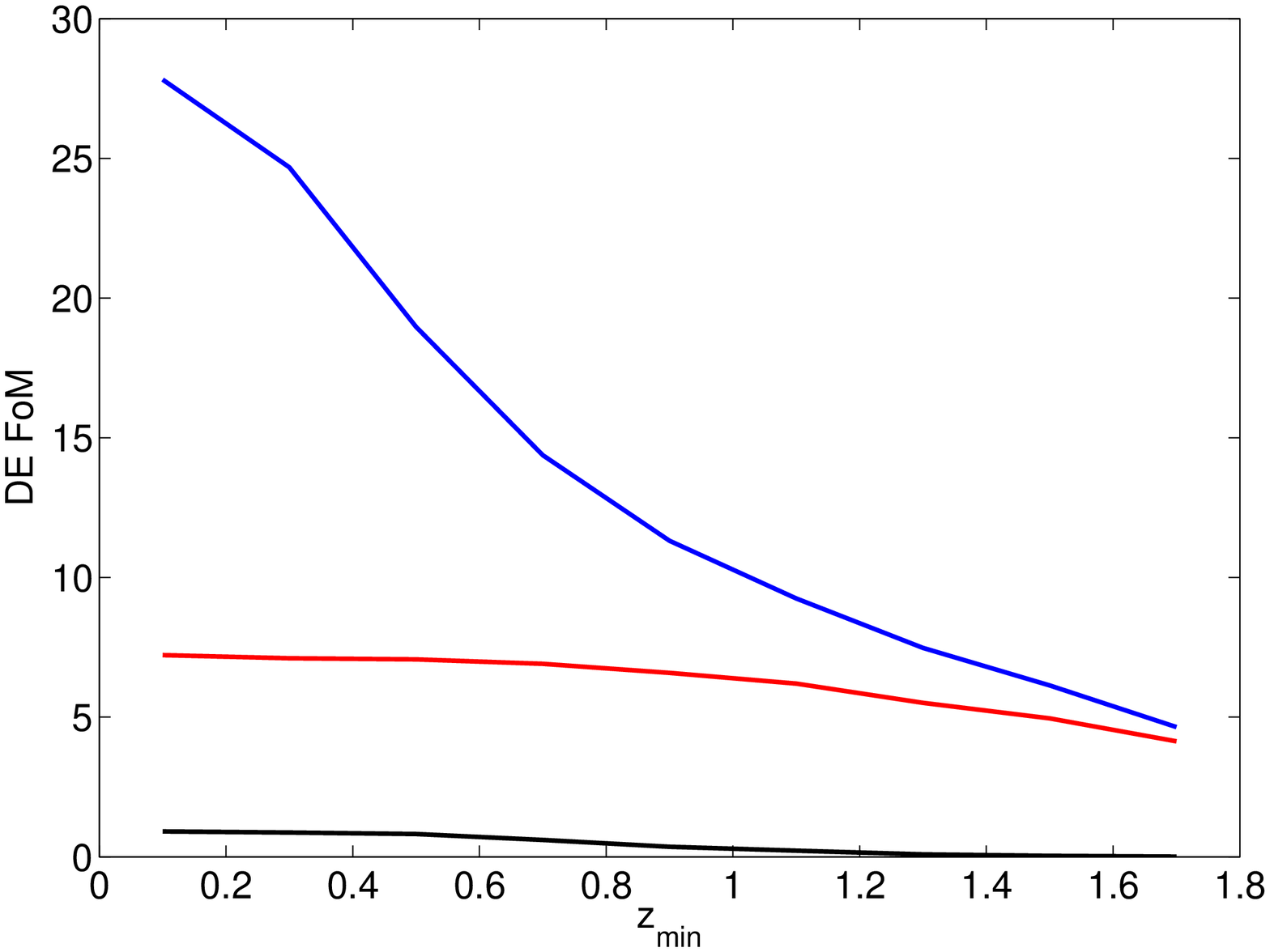}
       \includegraphics[width=3in,height=2in]{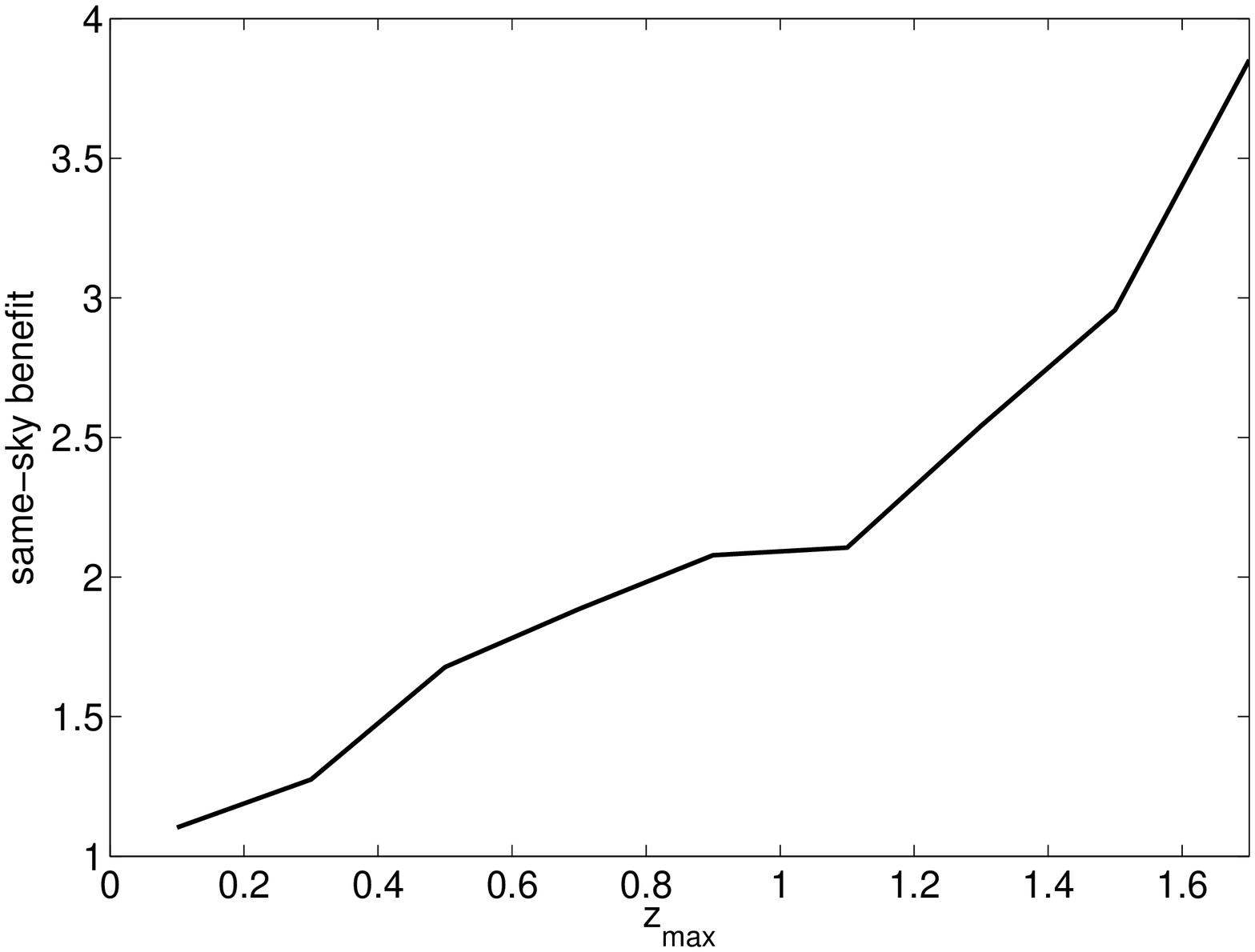}
       \includegraphics[width=3in,height=2in]{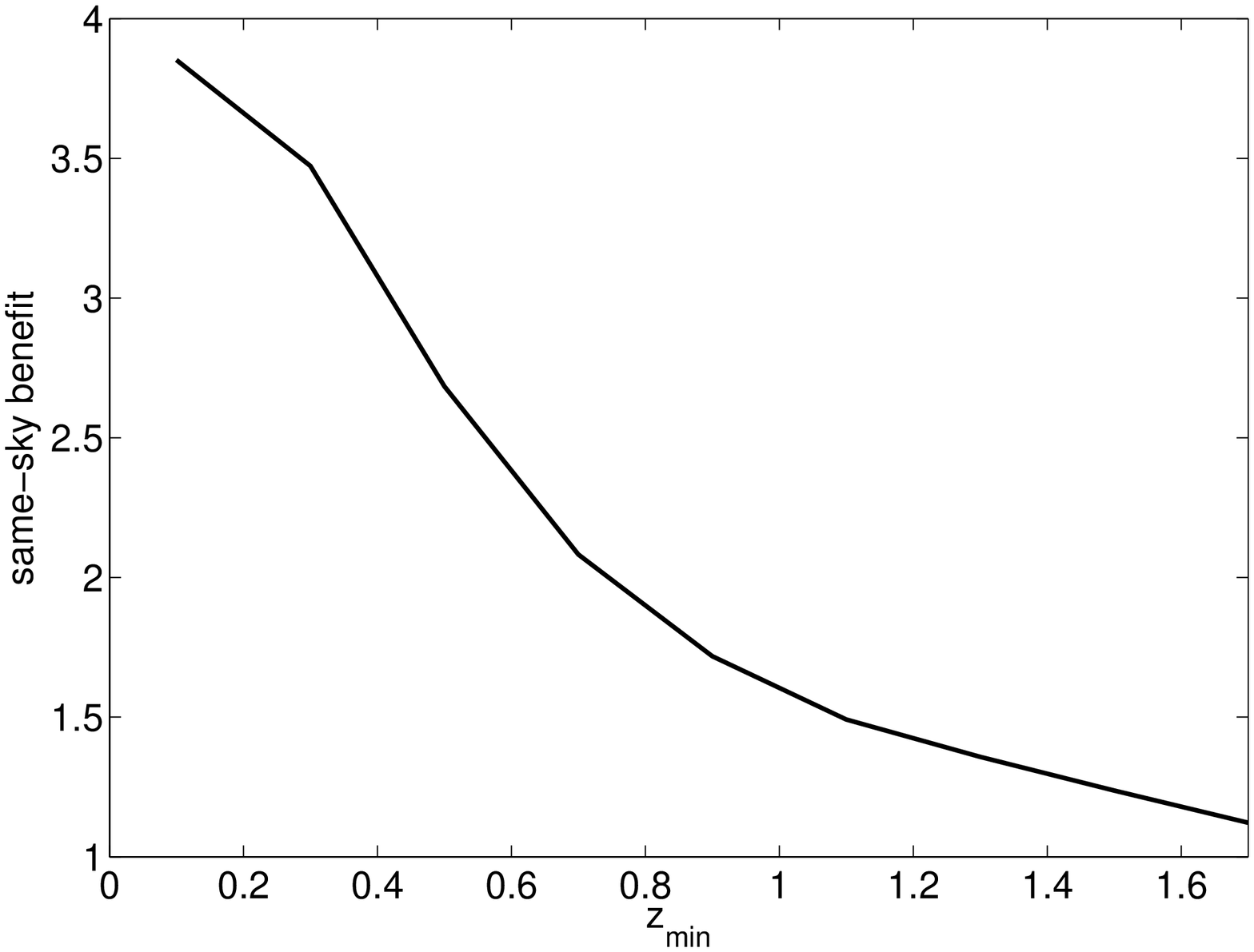}
\vspace{6mm}
\caption{This plot explores the importance of the z-coverage of our spec-z survey on individual and combined constraints. The spec-z survey $n(z)$ is assumed to be flat between its $z_{min}$ and $z_{min}$ values. All other survey assumptions are fixed at their fiducial values. [Left panels] Vary survey $z_{max}$ while keeping $z_{min}=0$ fixed. [Right panels] Vary $z_{min}$ while keeping $z_{max}$=1.7 fixed. [Top panels] DE FoM for spec-z $nn$ alone [black lines], the independent combination of spec-z and photo-z, $\epsilon\epsilon + nn$, [red lines] and the full joint combination including cross-correlations [blue lines]. [Bottom panels] Same-sky benefit, i.e. $FoM_{\epsilon\epsilon+n\epsilon+nn}/FoM_{\epsilon\epsilon+nn}$, as a function of $z_{min}$ and $z_{max}$.}
\label{fig:z_coverage}
  \end{flushleft}
\end{figure*}


We have investigated the benefit of overlapping our WGL \& LSS surveys in terms of shared area on the sky in section \ref{sec:partial_overlap}. In this section we examine the importance of survey overlap in the third dimension, along the line of sight. The importance of overlap in redshift space is intuitively very clear in our $C(l)$s formalism- each of our observables is the product of two window functions which can be thought of as particular z-space kernels. If the kernels are non-overlapping in $z$ then their product, and hence our observable, will be zero. It is important to note that while for our spec-z LSS survey the important quantity is $n^{i}(z)$, the redshift distribution of the target galaxies in a particular z-bin, $i$, for the photo-z WGL survey the relevant quantity is the lensing efficiency function for a particular bin which tends to peak at about half the peak redshift of the galaxy distribution.

Fig. \ref{fig:z_coverage} shows the impact of changing the fiducial z-range of our LSS survey, $z_{min} = 0 < z < z_{max} = 1.7$. We vary $z_{\textrm{min}}$ and $z_{\textrm max}$ for our spec-z survey independently. Redshift coverage of our photo-z survey remains fixed and we assume each survey covers the same 5000 deg$^2$. Maximising survey redshift coverage is clearly beneficial for all probe combinations. Including high-z regions is particularly important, with marked improvement for $z_{\textrm max} > 0.8$. This is to be expected as higher redshift coverage greatly increases survey volume.

As $z_{max}$ increases, same-sky benefit will obviously improve. The rate of improvement is steeper for $z>1.1$ suggesting that increased survey volume for the spec-z survey is the driving factor. The DES lensing kernal peaks below $z=1$ so the DES/DESI window function overlap is already ``locked in''. In contrast increasing $z_{min}$ sees the steepest fall in same-sky benefit for $z<0.8$. This demonstrates the importance of the spec-z $n(z)$ overlapping with the lensing kernel, losing this overlap greatly reduces the combined power of the surveys even if you retain much of the spec-z survey volume at high-z. 

The effect of z-coverage on MG constraints is broadly similar to that for DE. The only major difference is a bump in same-sky benefit as $z_{max}$ is increased from zero, peaking at $z_{max}\sim 0.5$, then dropping to $z_{max}\sim 0.7$ and rising slowly to $z_{max}\sim 1.7$. This suggests that the MG constraints benefit strongly from the $n\epsilon$ correlation at the peak overlapping redshifts of the WGL/LSS surveys. Achieving this overlap is at least equally important as the increased volume achieved by pushing to high redshift.

When designing overlapping spec-z and photo-z surveys a list of priorities is becoming apparent: good coverage of the lensing kernal by spec-z $n(z)$; joint coverage of a substantial fraction of the photo-z area and lastly a push to high-z to maximise spec-z survey volume. 

\subsection{LSS from the Photo-z Survey}
\label{sec:DES_LSS}

While this paper concentrates on the combination of WGL information from our photo-z survey with LSS information from our spec-z survey, it is worth discussing the fact that the photo-z survey obviously provides the information for a LSS analysis. In principle the spec-z survey could target galaxies for which we have shear estimates but the number density is so low and the lensing kernel so broad that their power as a WGL probe will be minimal.

A full analysis would include WGL and LSS from the photo-z survey plus LSS (inc. RSDs) from the spec-z and all their cross-correlations. We leave this complete analysis for a future paper but we have computed $\epsilon\epsilon$, $n\epsilon$ and $nn$ for the photo-z survey alone for comparison with our fiducial set-up in which the LSS information comes from the spec-z survey.

Naturally constraints from the WGL $\epsilon\epsilon$ alone are unchanged as we have always taken this probe from our photo-z survey. However we find that the $nn$ only constraints are much weaker for our photo-z survey due to the reduced sensitivity to RSDs from the broad tomographic bins. Combining $\epsilon\epsilon + nn$ does increase constraining power but by less, much less in the case of MG, than the photo-z + spec-z case because, without strong RSDs. It does not make sense to discuss a ``same-sky benefit'' in this case because we are dealing with datasets from the same photo-z survey, nevertheless we can say that there is strong improvement when the $n\epsilon$ correlations are included, producing a DE FoM nearly as strong as that from the photo-z + spec-z case. For MG the final constraint is much less strong, giving a MG FoM less than a quarter the size of that achieved by photo-z + spec-z. We can assert that, for DE, increased number density makes up for lower z-resolution/reduced RSD effects but the MG constraint suffers from the lack of RSDs which reduces the orthogonality of the WGL and LSS data.

In conclusion, the combination of our fiducial photo-z and spec-z surveys outperforms the joint WGL + LSS constraints from the photo-z survey alone, particularly in constraining deviations from GR. This is true even though our fiducial spec-z survey has shallower z-coverage and a significantly lower number density than our photo-z survey. While any complete analysis will exploit LSS information from the photo-z survey, where a suitable spec-z survey is available, joint constraints between surveys are strongly encouraged.

\subsection{Non-linear Scales}
\label{sec:kcuts}

\begin{figure}
  \begin{flushleft}
    \centering
       \includegraphics[width=3in,height=2in]{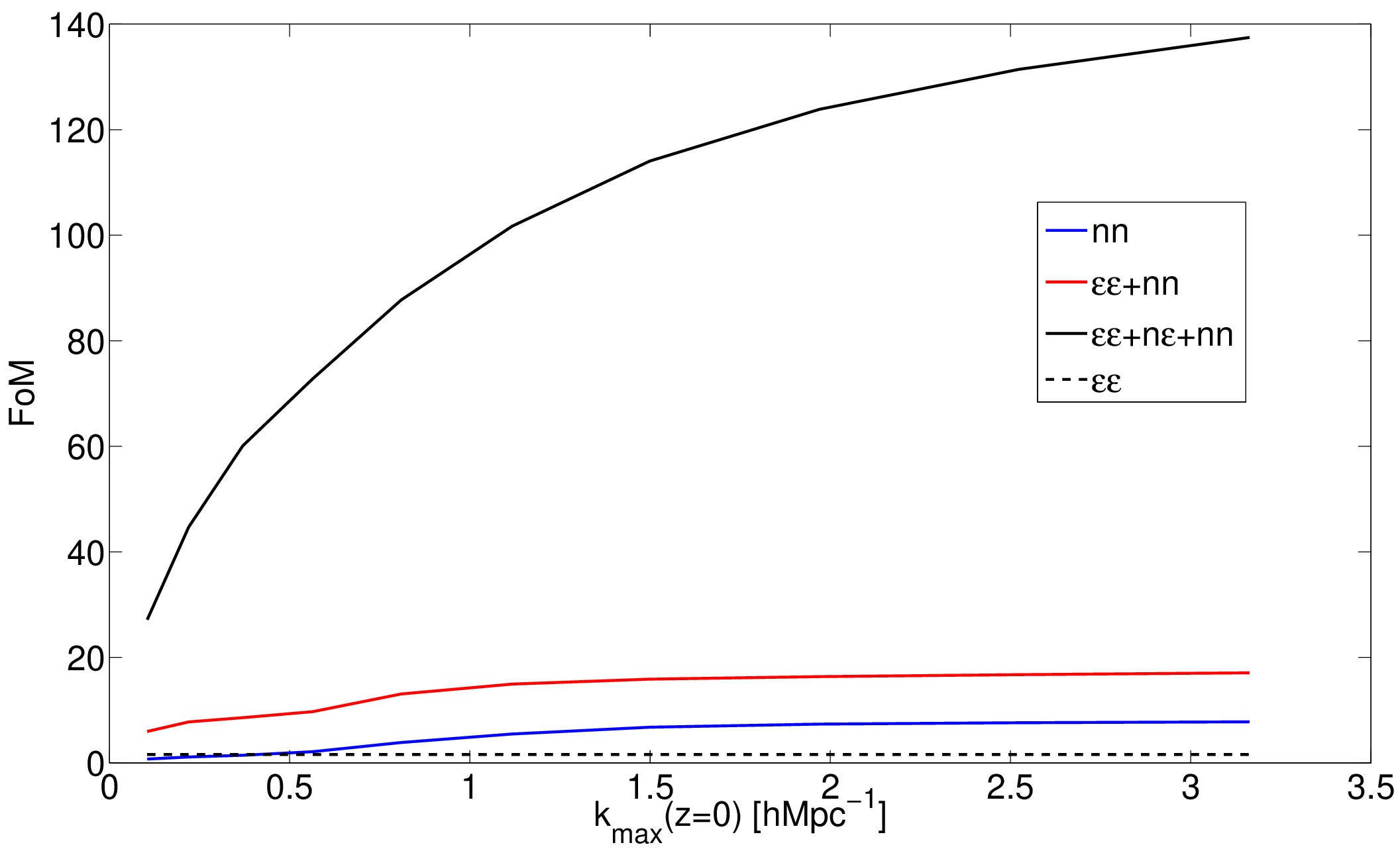}
       \includegraphics[width=3in,height=2in]{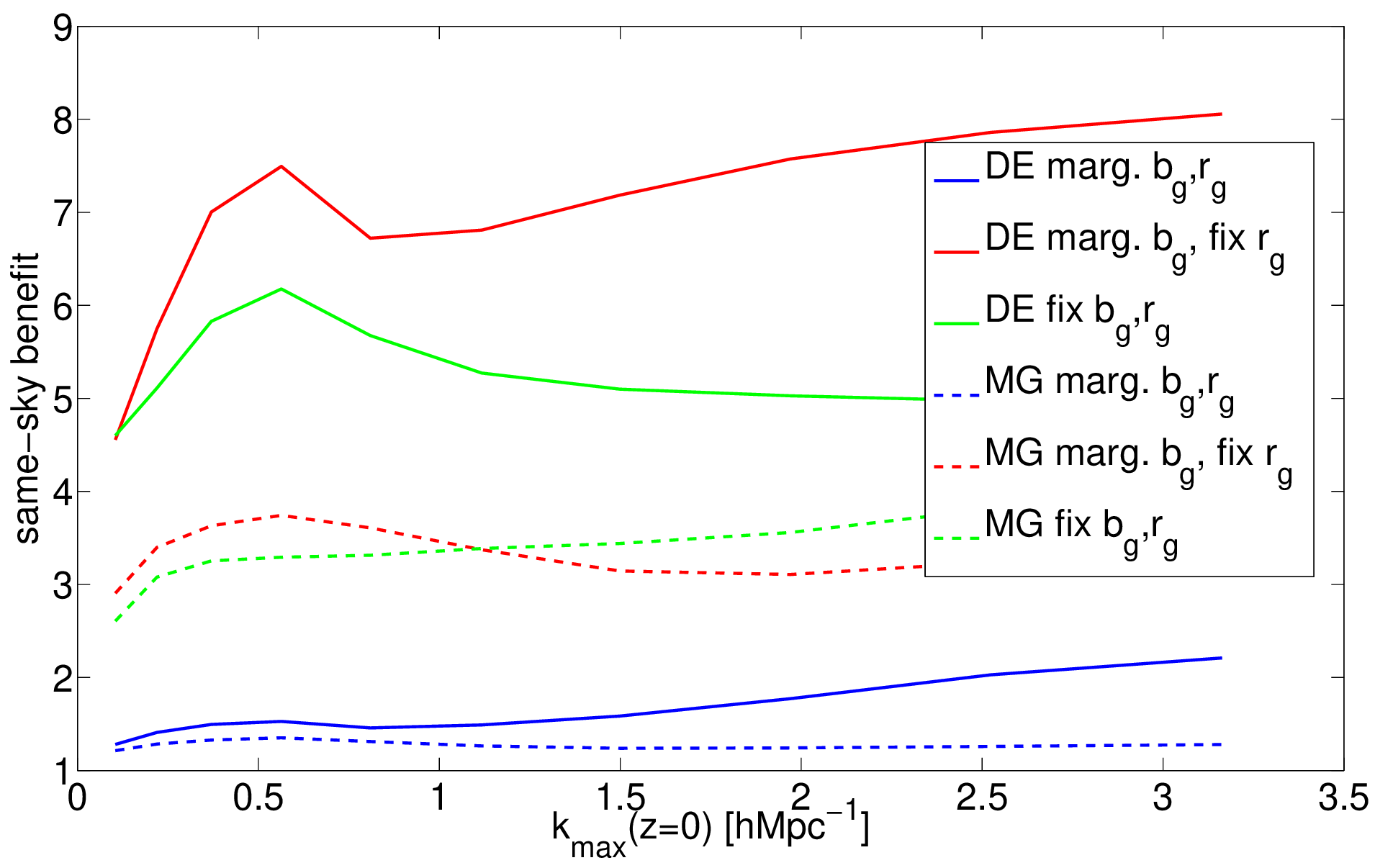}
\vspace{6mm}
\caption{This figure shows the impact of changing the scale at which we exclude non-linear clustering from our $nn$ and $n\epsilon$ analysis. The x-axis is the maximum k value included (at z=0). We alter the k-cuts according to the recipe of \citet{Rassatea_2008} where we vary the X parameter away from its fiducial value of 0.2, equivalent to $k\sim0.25hMpc^{-1}$ on this plot. [Upper panel] FoM as a function of the maximum k-mode included in the spec-z analysis for $nn$ [black], $\epsilon\epsilon+nn$ [red] and $\epsilon\epsilon + n\epsilon + nn$ [blue]. $\epsilon\epsilon$ [black dash] FoM is shown for comparison, it is never subjected to a k-cut. [Lower panel] Same-sky benefit, i.e. $FoM_{\epsilon\epsilon+n\epsilon+nn}/FoM_{\epsilon\epsilon+nn}$, as a function of maximum k-mode included in LSS analysis. Same-sky benefit is shown for DE [solid] and MG [dashed] for the cases where both $b_g$ and $r_g$ are marginalised over [blue], $b_g$ is marginalised over but $r_g$ is fixed [red] and both $b_g$ and $r_g$ are held fixed [green].}
\label{fig:kcut}
  \end{flushleft}
\end{figure}

In the preceding sections we have assumed a cut which removes modes from our LSS analysis corresponding to non-linear and quasi-linear scales. The prescription we use is taken from \citet{Rassatea_2008}, section 4.3. This approach removes scales smaller than $l_{max} = k_{max}\chi(z_{med}^{i})$, where $\chi(z_{med}^{i})$ is the comoving distance of the median redshift of tomographic bin $i$, i.e. the scale-cut is redshift dependent by bin. $k_{\textrm max}$ is defined by considering only scales for which $\sigma(R) < X$ where 
\begin{equation}
\sigma^{2}(R) = \int \Delta^{2}(k) \frac{dk}{k} \frac{9}{(kR)^{6}} \left[ \textrm{sin}(kR) - kR \textrm{cos}(kR) \right]^{2}. 
\end{equation}
\citet{peacockd96} eqn. 42 defines a $R$ value and $k = 2\pi / R$ relates this to a $k_{\textrm max}$. Our fiducial choice is $X = 0.2$, corresponding to $k_{\textrm max} \sim 0.25 h Mpc^{-1}$.  
Figure \ref{fig:kcut} shows the impact of changing this assumption. We show the change in FoMs and same-sky benefit from varying the $X$ value which, in turn, changes the maximum $k$-value which we include in our $nn$ and $n\epsilon$ $C(l)$s.


As expected, including more non-linear scales in our analysis improves FoM from all forecasts which include LSS observables. For $nn$ alone, when we include smaller scales, the $nn$ survey quickly overtakes the constraining power of the $\epsilon\epsilon$ survey (shown as a horizontal line for reference, it never gets cut on ell), suggesting that our NL-cut exerts a strong constraint on the LSS observable. We see similar improvement for the $\epsilon\epsilon + nn$ combination, with both exhibiting plateaus above $k\sim 1-1.5hMpc^{-1}$. In contrast the full combination of WGL \& LSS surveys, including cross-correlations, $\epsilon\epsilon + n\epsilon + nn$ shows a very strong improvement in FoM right down to $k=3hMpc^{-1}$. In this case increasing $k_{max}(z=0)$ from 0.25$hMpc^{-1}$ to 3$hMpc^{-1}$ improves FoM by a factor of four. We would not have necessarily expected this behaviour but it seems to suggest that the inclusion of highly nonlinear scales combined with the WGL/LSS cross-correlation breaks cosmological parameter degeneracies more fundamental than the simple control of $b_g$. It could simply be due to the fact that the cross-correlation combination has another set of nuisance parameters to constrain ($r_g$) that continue to benefit from NL scales, even after $nn$ itself has exhausted its constraining power. Indeed when we fixboth $b_g$ and $r_g$, the $\epsilon\epsilon + n\epsilon + nn$ constraint increases strongly with large $k_{max}$ but quickly plateaus above $k_{max}>1$, while $nn$ and $\epsilon\epsilon+nn$ continue to increase. 

One constant in all our results is the rapid increase in the constraining power of $\epsilon\epsilon + n\epsilon + nn$ as we increase our fiducial $k_{max}$, leading to a rapid improvement in same-sky benefit. We are content then that our choice of fiducial $k_{max}(z=0)\sim0.25$ is a sensibly conservative one. Improvements in our understanding of non-linear galaxy and DM clustering are always to be welcomed and may improve the constraining power of our LSS observables but they are not essential to see strong improvements in constraining power from the overlap of spec-z and photo-z surveys.

\section{Discussion}
\label{sec:discussion}

This paper, like others before it \citep{bernstein_cai_2011,gaztanaga_pau_2012,cai_bernstein_2012,duncanea_2013}, has demonstrated the benefits of combining a photometric WGL survey with a spectroscopic LSS survey. When their independent likelihoods are simply added we see more than a factor of four improvement in DE FoM on the best either can do alone. This improvement is particularly marked in a two parameter modified gravity model where constraints from each probe are orthogonal and their combination breaks an important degeneracy, here the $\epsilon\epsilon + nn$ combination performs more than two orders of magnitude better than either probe alone. It is worth noting that this very strong improvement is only visible when one goes beyond the common $\gamma$ parameterisation of modified linear growth to a two (or more) parameter modified gravity model which is sensitive to the fact that the WGL and LSS observables are sensitive to different combinations of the metric potentials.

Beyond this we have used a simple combined probes formalism based on projected angular power spectra to calculate the cross-power spectra between our photo-z and spec-z surveys and also their full joint covariance matrix. The inclusion of these cross-correlations models the extra data available when we have both of these cosmic probes observed on the same patch of sky. Having overlapping surveys of this nature provides a range of benefits. Photometry is necessary to construct a target list for spectroscopy, while spectroscopic data can help calibrate the photometric redshift distribution. Systematic effects such as galaxy bias or Intrinsic Alignments are often more accurately characterised from multiple overlapping datasets. In this paper we have concentrated on the improved constraints on cosmology and nuisance parameters that come from conducting a full joint likelihood calculation in our $C(l)$s formalism.

This has allowed us to define a same-sky benefit factor- the improvement when these $n\epsilon$ cross-correlations are included. Four our fiducial forecast assumptions we see strong positive same-sky improvements of nearly a factor of four for DE and more than a factor of two for MG.

Any such forecast is a complicated calculation, within which many assumptions are made which can radically affect the final results. We have tried to methodically disentangle a number of the most important assumptions in an effort to quantify their impact and produce the most robust range of forecasts possible.

We are confident that the general trend of our fiducial survey results are robust to the inclusion of priors from a Planck-like CMB experiment and for the type of feature-full spec-z $n(z)$ produced by any specific spectrograph/telescope combination, target selection choices and survey strategy. These choices are investigated in more detail in our companion paper \citet{jouvelea_2013}.

The importance of galaxy-shear cross correlations for controlling galaxy bias has been extensively noted \citep{yoo_seljak_2012,deputter_dore_das_2013,asoreyea_2013}. We show that indeed choices of galaxy bias model, nuisance parameterisation and galaxy population bias amplitude can all significantly effect both LSS-only constraints and combined WGL+LSS constraints. Galaxy bias modelling is an area of active research interest which will benefit greatly from improved observations over the coming years. Currently the state of our knowledge of $b_g$ is limited enough the necessitate the inclusion of nuisance parameters which, when marginalised over, are a way of including our uncertainty about the true galaxy bias into a forecast. More nuisance parameters decreases our ability to measure cosmology but will produce a more robust, less biased final result. We remove truly non-linear scales (for which $b_g$ modelling is particularly uncertain) from our analysis entirely with judicious cuts on small scales.

We show that, while increasing the number of galaxy bias nuisance parameters does reduce our constraining power, there is little decrease beyond a $4 \times 4$ grid of nuisance parameters in k/z space, i.e. a bias model with 17 free parameters. The same-sky benefit does increase with more uncertainty in galaxy bias, supporting the assertion that the $n\epsilon$ correlation can control for $b_g$, however, even if we assume bias is known perfectly there is a factor of three benefit from the extra correlations offered by overlapping surveys.  

A similar effect is observed when we increase the uncertainty in the photometric redshift error. As the $n\epsilon$ correlation can go some way towards ``calibrating'' this error, there is more scope for improvement when the photo-z error is less well understood and the same-sky benefit is correspondingly higher. However, we want to emphasis that we find, as with galaxy bias, that there is still substantial improvement due to $n\epsilon$ cross-correlations even in the case where the photo-z error is assumed to be perfectly described.

The forecasting assumption that most impacts the same-sky benefit is our assumed knowledge of the galaxy-shear cross-correlation coefficient, $r_g$. If we allow this to vary with the same freedom as our fiducial $b_g$ model then same-sky benefit is reduced to less than a factor of 1.2. However there are strong theoretical arguments that suggest $r_g$ is close to unity, at least on the linear and quasi-linear scales we include here. We suggest that the very low same-sky benefits found from aggressive marginalisation over $r_g$ are overly pessimistic \citep{gaztanaga_pau_2012}.

As well as assessing the difference between combined constraints from surveys on different parts of the sky versus surveys which completely overlap, we also look at the effects of partial overlap, both in area on the sky and in z-coverage. We clearly see that even a partial overlap is beneficial, particularly for MG, where half the full 5000 deg$^2$ overlap benefit comes from the first 1000 deg$^2$. Our z-coverage analysis shows two complementary sources of improvement. Most important is that the spec-z survey covers the peak of the WGL lensing kernel. Once this requirement has been met, pushing to higher redshift and thus increased volume for the spec-z survey continues to be very beneficial. 

On the issue of same-sky improvement, the benefit from the extra cross-correlations available when our LSS and WGL surveys overlap on the sky, we can see a range of results depending on a variety of assumptions that are made when making forecasts. Clearly the worst same-sky benefit results come from aggressively marginalising an unknown galaxy-shear cross-correlation, $r_g$, which results in a same-sky benefit factor of 1.2. This is highly pessimistic and there are strong arguments, both theoretical \citep{gaztanaga_pau_2012} and observational \citep{comparat_bias_2013}, that $r_g$ is very close to unity on the linear scales for which we consider LSS data. 

For some assumptions we see very strong same-sky improvements. Most promisingly the example we take of a $n(z)$ based on a specific target selection and survey strategy scenario shows a DE same-sky benefit of more than a factor of ten. While this number will be very dependent on the details of target selection, survey strategy etc, it is still very promising when we consider the application of this analysis to real survey data. In addition, the targetting of more strongly biased galaxy populations, not only makes the LSS probes more constraining but increases the same-sky factor. This is clearly of relevance to the \citet{mcdonald_seljak_2009} technique for the control of cosmic variance and we intend to produce a more comprehensive analysis in a future paper.

In general we see that DE and MG follow very similar trends as we perturb our assumptions away from the fiducial model. The major difference remains the fact that MG benefits so strongly from the independent combination of WGL and LSS due to the orthogonality of the constraint contours. This produces such a dramatic improvement that the same-sky benefit is correspondingly less pronounced than in the DE case. We see lower same-sky benefits for MG than for DE for all forecast assumptions (with the exception of the case without RSDs which is only included to demonstrate the power of the use of RSDs). Nevertheless the MG same-sky benefit remains at the level of a factor of two or more for most sensible forecast assumptions.



There are a range of possible extensions to the work we present in this paper. The joint analysis of different cosmological data sets is becoming more ambitious. We hope to extend our $C(l)$s formalism to allow the cross-correlation of an arbitrarily large number of observables in any bin combinations. Among other things this would allow us to include LSS information from our photo-z survey as well as break our galaxy populations up into population samples which are differently biased. \citet{mcdonald_seljak_2009} suggest that this is an effective way to reduce cosmic variance. We also aim to include cosmic magnification in our future efforts as well as conduct a more detailed study of the trade-off in accuracy due to projection effects when modelling a spec-z galaxy survey in projected angular power spectra.

\section{Conclusions}
\label{sec:conclusions}

Joint survey analysis is clearly an essential part of cosmology if we are to make the most of the unprecedented data sets shortly to become available from a range of cosmic probes. Different probes go beyond the sum of their parts through the breaking of degeneracies between cosmological parameters. In addition they can help to constrain systematic effects and unknown physical quantities e.g. halo masses or cluster-mass relations. Multiple probes on the same patch of sky often observe the same objects and there is obvious synergy between photometric and spectroscopic surveys in terms of target selection and photo-z error calibration.

We have produced a range of joint forecasts, combining generic photometric WGL and spectroscopic LSS surveys. Even a simple forecast of this kind requires a large number of assumptions, often implicit. We have tried to lay bare every part of the process and conduct a sensitivity analysis by varying each assumption in turn and quantifying their impact on individual and combined constraints relative to eachother.

Throughout our sensitivity analysis we see some constant trends: (i) the combination of WGL from a photo-z survey and LSS from a spec-z survey greatly improve our ability to measure the equation of state of DE and deviations from General Relativity, by a factor of four in DE FoM for our fiducial surveys compared to photo-z WGL alone, (ii) in the MG case in particular the orthogonal nature of the constraints from both probes produces a very strong joint constraint once degeneracies in the MG parameters are broken, improving our MG FoM by over two orders of magnitude for non-overlapping surveys compared to photo-z WGL alone, and (iii) there is a significant benefit from overlapping surveys on the same patch of sky which allows us access to the $n\epsilon$ cross correlation between probes and the full covariance matrix including all off-diagonal elements, giving an extra factor of four for DE and more than two for MG compared to non-overlapping surveys with our fiducial assumptions.

Different groups have produced conflicting results on the question of the same-sky benefit from overlapping surveys such as those we consider. While \citet{gaztanaga_pau_2012} see significant improvement from same-sky overlap, \citet{cai_bernstein_2012} predict little improvement over the independent combination of WGL/LSS as if they were on different patches of sky. Our results are more in agreement with the findings of \citet{gaztanaga_pau_2012} as we see good same-sky benefit from most sensible survey forecast assumptions. One possible source of this disagreement is the relative constraining power of the LSS observable alone. We have been relatively conservative in exclusion of quasi-linear and non-linear scales and also in a relatively aggressive marginalisation over galaxy bias. It is possible that, if we change these assumptions and allow the LSS survey alone to be more constraining, then the benefit from combining with WGL or the $n\epsilon$ cross-correlation would be correspondingly diminished. We perhaps see some hints of this as we increase our $k_{max}$ while assuming $b_g$ and $r_g$ are exactly known. In addition the reduction of cosmic variance from an implementation of the \citet{mcdonald_seljak_2009} technique may improve the spec-z survey's constraining power and reduce the impact of same-sky cross-correlations.

What is clear throughout the literature is that combined probes of the kind available now and in the coming years can measure cosmology to very high precision. We have presented one flexible framework for this type of joint constraint (based of course on previous work \citep{bernstein_2008,joachimi_bridle_2009}. Even in the simple scenario here, where have only considered one observable from each of our two surveys, it is clear that the range of assumptions that go into the forecast make the prediction a complex one. As the number of observables included in a simultaneous joint analysis are increased this effect will only become more pronounced. It is vitally important that all assumptions are stated explicitly and examined in isolation to determine their effect relative to others. If we conduct this process correctly the prize is enormous: highly precise cosmological measurements, far beyond anything available to probes considered in isolation. Only in this way will we be fully able to exploit our available data and more precision cosmology onto the next level. 

\section*{Acknowledgements}

OL acknowledges a Royal Society Wolfson Research Merit Award, a Leverhulme Senior Research Fellowship and an Advanced Grant from the European Research Council.
FBA thanks the Royal Society for support via an URF.
SB acknowledges support from European Research Council in the form of a Starting Grant with number 240672.

Many thanks to all those who engaged in very helpful discussions on the combined probes and same sky issues, especially Enrique Gaztanaga, Jacobo Asorey, Gary Bernstein and Yan-Chuan Cai.
We are indebted to the DESpec collaboration for useful discussion and collaboration.
Thanks to Jochen Weller for supplying the Planck FM and Tom Kitching on its relation to Planck data.
Thank you to Anais Rassat, Ole Host, Lisa Voigt and Lucy Clerkin for insights into nuisance parameters and forecast assumptions.

\bsp

\bibliographystyle{mn2e}
\bibliography{bibliography_DK}

\label{lastpage}

\end{document}